\documentclass[twocolumn]{pasj00}
\usepackage[varg]{txfonts}
\color{black}

\newcommand{\fluxunit}{erg~cm$^{-2}$~s$^{-1}$}

\begin{document}
\Received{$\langle$reception date$\rangle$}
\Accepted{$\langle$accepted date$\rangle$}
\Published{$\langle$publication date$\rangle$}
\SetRunningHead{H. Akamatsu et al.}{Shock Front in A~3376 with Suzaku}
\title{
X-Ray View of the Shock Front in the Merging Cluster Abell 3376 with Suzaku\thanks{Last update: \today}}
\author{
 Hiroki {\sc Akamatsu},\altaffilmark{1}
 Motokazu {\sc Takizawa},\altaffilmark{2}
  Kazuhiro {\sc Nakazawa},\altaffilmark{3}
 Yasushi {\sc Fukazawa},\altaffilmark{4} \\
 Yoshitaka {\sc Ishisaki},\altaffilmark{1}
 and Takaya {\sc Ohashi}\altaffilmark{1}
}
\altaffiltext{1}{
  Department of Physics, Tokyo Metropolitan University, 1-1
  Minami-Osawa, Hachioji, Tokyo 192-0397}
\altaffiltext{2}{
   Department of Physics, Yamagata University, Kojirakawa-machi
   1-4-12, Yamagata 990-8560}
\altaffiltext{3}{
  Department of Physics, The University of Tokyo, 7-3-1 Hongo,
  Bunkyo-ku, Tokyo 113-0033}
\altaffiltext{4}{
   Department of Physical Science, Hiroshima University, 1-3-1
   Kagamiyama, Higashi-hiroshima, Hiroshima 739-8526}
\email{h\_aka@phys.se.tmu.ac.jp}
\maketitle
\begin{abstract}
  We report on a Suzaku measurement of the shock feature associated
  with the western radio relic in the merging cluster A~3376\@.  The
  temperature profile is characterized by an almost flat radial shape
  with $kT \sim 4$ keV within $0.5 r_{200}$ and 
a rise by about 1 keV inside the radio relic. Across the relic
    region ($0.6-0.8 r_{200}$), the temperature shows a remarkable
    drop from about 4.7 keV to 1.3 keV\@. 
    This is a clear evidence that the radio relic really corresponds to a shock front possibly
  caused by a past major merger.  The observed sharp changes of the
  temperature and electron density indicate the Mach number ${\cal M} \sim 3$.  
  The radial entropy profile is flatter than the prediction ($r^{1.1}$) of numerical simulations within $0.5 r_{200}$, 
and becomes steeper around the relic region. 
These observed features and time-scale estimation consistently imply that the ICM around the radio relic has 
experienced a merger shock and is in the middle of the process of dynamical and thermal relaxation.

\end{abstract}
\section{Introduction}
\label{sec:intro}
Clusters of galaxies are believed to grow through gas accretion from
large-scale filaments and mergers of subclusters.  Cluster mergers are
the most energetic events in the Universe after the Big Bang with the
total kinetic energy of the colliding subclusters reaching $10^{65}$
ergs.  The kinetic energy is converted to thermal energy by driving shocks and turbulence.  
Existence of mega-parsec scale radio emission (halos and relics) in merging clusters
  indicates that those shocks and turbulence operate as main
  mechanisms of particle acceleration (see \cite{ferrari08} and
  reference therein).  
Radio relics are considered to be the synchrotron emission generated through the interaction
 between relativistic electrons, accelerated by a shock, and amplified magnetic field in the intracluster medium (ICM)\@.
  Although these radio features provide direct evidence of particle
  acceleration, the detection of non-thermal (hard) X-ray emission
  caused by relativistic particles is still controversial \citep{nevalainen04,ajello09}.  
  Currently, hard X-ray observations set lower limits of magnetic fields in ICM~\citep{wik09,sugawara09, clarke06}.  
  Based on recent studies, it is believed that magnetic field of 0.1--10 $\mu$G exists in ICM\@.


Theoretical studies of cluster merger shocks provide rich information
about the magnetic field, turbulence, particle acceleration and their
time evolution (see e.g.\ \cite{brueggen11} for a review). Merger
shocks are characterized by a Mach number ${\cal M} \lesssim 3$,
causing amplification of magnetic fields to $B \sim 10\ \mu$G and
production of relativistic particles \citep{takizawa00,takizawa08}.
Also, turbulence is considered to give a significant pressure support
in the outer regions, with around 10\% of the thermal pressure at
least for a few Gyr after a major merger \citep{parrish11}.  Recent
numerical simulations include interaction of merger shocks with the
filamentary cosmic web structures outside of clusters, and the results
indicate that shock fronts are more enhanced in the filament direction
\citep{paul11}.

Though past X-ray observations have shown images and temperature
structures of cluster mergers, there are only few clusters for which
clear evidence of the shocks has been obtained (1E0657-56, A520, and
A2146: \cite{markevitch05, clowe06,russell10}). The low X-ray surface
brightness in the outer regions has hampered precise measurements of
gas temperature and density in the pre-shock region in particular.
Recently, existence of shock fronts with a Mach number of about 2 was
confirmed in the radio relic region of A~3667
\citep{finoguenov10,akamatsu11b}.  Additionally, a remarkable radio
relic indicated a shock front with a Mach number of 4.6 based on the
measurement of radio spectral index (CIZA2242: \cite{weeren10}).
Thus, radio relics are the good probe to identify merger shocks and
combination with X-ray data will yield important information about the
gas dynamics associated with cluster mergers.  Since the radio relics
are mainly found in the outskirts of clusters where the gas density is
low, sensitive X-ray observations are needed, especially in the
upstream side of the shocks.  The X-ray data will allow us to estimate
the parameters of the gas and the magnetic field, and enable us to
look into the actual process occurring during the cluster evolution.

In this paper, we present a new X-ray evidence of the
  shock associated with the radio relic in Abell 3376, a well-known
  merging cluster with irregular morphology at $z=0.046$.
Previous studies showed the global mean temperature to be 4.0 keV and the
existence of a pair of Mpc-scale radio relics revealed by 1.4 GHz VLA
NVSS observations~\citep{bagchi06}.  Another feature of A~3376 is the
report of the hard X-ray signal with
BeppoSAX/PDS~\citep{nevalainen04}.
However, Suzaku HXD observation gave an upper limit which did not
exclude the BeppoSAX flux \citep{kawano09}.  We need to await further
sensitive observations in the hard X-ray band.

We use $H_0 = 70$ km s$^{-1}$ Mpc$^{-1}$, $\Omega_{\rm M}=0.27$ and
$\Omega_\Lambda = 0.73$, respectively, which give 54 kpc per arcminute
at $z = 0.046$.  The virial radius is approximated by $r_{200} = 2.77
h_{70}^{-1} (\langle T\rangle /10 {\rm keV})^{1/2}/E(z)\ {\rm Mpc} $,
where $E(z)=(\Omega_{\rm M}(1+z)^{3}+1-\Omega_{\rm M})^{1/2}$
\citep{henry09}.  For our cosmology and redshift, $r_{200}$ is 1.86
Mpc ($= \timeform{34.6'}$) with $kT = 4.7$ keV\@.  We employ solar
abundance defined by \citet{anders89} and Galactic absorption with
$N_{\rm H}=5.8 \times10^{20}$cm$^{-2}$~\citep{dickey90}.  Unless
otherwise stated, the errors correspond to 90\% confidence for a
single parameter.

\begin{table*}[t]
\label{tab:obslog}
\begin{center}
\caption{Suzaku observations of A~3376 and the background region.}
\begin{tabular}{cccccccccc}
\hline
Target (Obs.\ ID) & Position &Start time &Exposure$^\ast$ &Exposure$^\dagger$ \\
  & (R.A., Decl.)   & (UT) & (ks) & (ks) \\
\hline
A~3376c (100034010) & 
   $(90.56, -39.94)$ & 2005/10/06 14:46:08 & 110.4 & 78.0 \\ 
A~3376w (800011010) & 
   $(90.05, -39.98)$ & 2005/11/07 14:15:05 & 120.0 & 99.9 \\
Q~0551-3637 (703036020) & 
   $(88.19, -36.63)$ & 2008/05/14 13:25:55 & 18.8 & 15.3\\
\hline
$\ast$: no selection with COR2 &\multicolumn{2}{l}{ ${\dagger}$: COR2 $> 8$ GV}\\\
\end{tabular}
\end{center}
\end{table*}

\begin{figure}[t]
\begin{center}
 \includegraphics[scale=0.45]{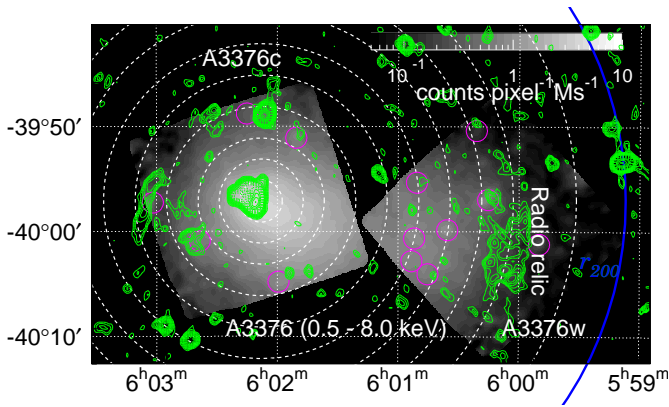}
\end{center}
\caption{ X-ray image of A~3376 in the energy band 0.5-8.0 keV, after
  subtraction of the NXB with no vignetting correction and after
  smoothing by a 2-dimensional Gaussian with $\sigma =16$ pixel
  =$\timeform{17''}$.  The large blue circle shows the virial radius
  of A~3376 with $r=34'.6$, and white dotted circles show the annular
  regions used for the spectral analysis.  Small magenta circles show
  point sources which are detected by {\it wavdetect} (see text).  The
  VLA 1.4 GHz radio image is shown with green contours.  }
\label{fig:suzaku_image}
\end{figure}

\section{Observations and Data Reduction}

As shown in Fig.\ \ref{fig:suzaku_image}, Suzaku carried out two
pointing observations of Abell 3376 along the merger axis in October
and November 2005, designated as A~3376c, A~3376w for the center and the
west relic.  The observation log is summarized in
Table~\ref{tab:obslog}.  All the observations were performed with
either normal $5\times5$ or $3\times3$ clocking mode.  The combined
observed field extends to the virial radius of A~3376 ($34'.6 \sim
1.86$ Mpc) indicated with blue circle in Fig.\ \ref{fig:suzaku_image}.

The XIS instrument consists of 4 CCD chips: one back-illuminated (BI:
XIS1) and three front-illuminated (FI: XIS0, XIS2, XIS3) ones.  Even
A~3376 observations were carried out only 3--4 months after the launch,
the IR/UV blocking filters had some contamination accumulated at the
time of the observations.  We included its effect and uncertainty on
the soft X-ray effective area in our analysis.  We used HEAsoft
version 6.11 and CALDB 2011-06-30 for all the Suzaku data analysis
presented here.  We performed event screening with the cosmic-ray
cut-off rigidity (COR) $> 8$ GV to increase the signal to noise ratio.
We extracted pulse-height spectra in 10 annular regions whose boundary
radii were $~\timeform{2'}, ~\timeform{4'}, ~\timeform{6'},
~\timeform{9'}, ~\timeform{12'}, ~\timeform{15'}, ~\timeform{18'},
~\timeform{21'}, ~\timeform{24'}, \timeform{27'}~\rm
and~\timeform{31'}$, with the center at ($\timeform{06h01m00s},
-39^\circ57'07''$).  
 We analyzed the spectra in  the 0.5--10~keV range for the FI detectors and 0.5--8~keV for the BI
  detector.  The energy range of 1.7-1.9 keV was ignored in the
  XIS spectral fitting, because the response matrix around Si-K
  edge had some residual
  uncertainties\footnote{http://heasarc.nasa.gov/docs/suzaku/analysis/sical.html}
  .  In all annuli, positions of the calibration sources were masked
  out using the {\it calmask} calibration database (CALDB) file.

\begin{figure}[t]
\begin{center}
 \includegraphics[scale=0.3,angle=-90]{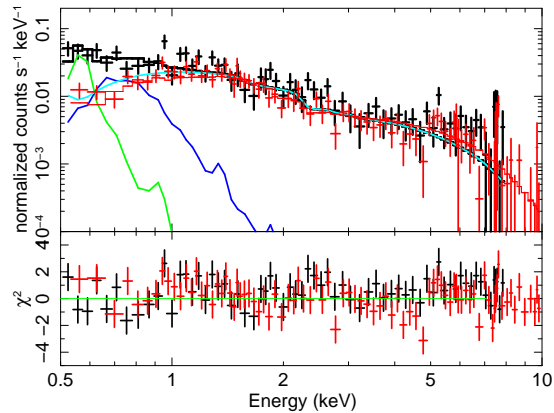}
\end{center}
\caption{ The spectrum of the Q0551 field used for the background
  estimation, after subtraction of NXB and the source.  The XIS BI
  (black) and FI (red) spectra are fitted with CXB + Galactic
  components (LHB, MWH) ({\it apec+wabs(apec+powerlaw)}).  The CXB
  spectrum is shown with a black curve, and the LHB and MWH components
  are indicated by green and blue curves, respectively.  }
\label{fig:bgd}
\end{figure}

\section{Spectral Fitting}\label{sec:spec}
\subsection{Method of Spectral Fits}\label{sec:spec_model}

In the analysis of A~3376 data, we followed our
  previous analysis for the Suzaku observation of A~3667
  ~\citep{akamatsu11b}, which shared many similar features 
  but with much brighter surface brightness than the present object.  
The basic method of the spectral fit is described in subsection 4.1 of
\citet{akamatsu11b}.  The observed spectrum was assumed to consist of
thin thermal plasma emission from the ICM, the local hot bubble (LHB)
and the milky way halo (MWH) as the Galactic foreground components, cosmic
X-ray background (CXB) and non-X-ray background (NXB)\@.  The NXB
component was estimated from the dark Earth database by the ${\it
  xisnxbgen}$ FTOOLS \citep{tawa08} and was subtracted from the data
before the spectral fit.  To adjust for the long-term variation of the
XIS background due to radiation damage, we accumulated the NXB data
for the period between 50 days before till 150 days after the
observation of A~3376\@.

The spectrum of the Galactic background emission can be represented by two thin-thermal plasma models, 
by {\it apec} with 1 solar abundance.  
The estimation of the Galactic component and CXB is described in subsection \ref{sec:bgd}.  
Because the Galactic and CXB components have almost uniform 
spatial
 distribution in the XIS field of view, 
we generate uniform auxiliary response files (ARF) for those emission.  
As for the ICM emission, we generated the ARF assuming the $\beta$-model surface brightness with
$\beta=0.40$, and $r_{c}=\timeform{2.03'}$ \citep{cavagnolo09} using {\it xissimarfgen} \citep{ishisaki07}.

For the spectral fits, we used XSPEC ver12.7.0.  
We carried out spectral fit to the pulse-height spectrum in each annular region separately.  
In the simultaneous fit of the BI and FI data, only the normalizations were allowed to be different between them, 
although we found that the derived normalizations were quite consistent between the two within 15\%. 
In the central regions within $\timeform{25.2'}$, free parameters were temperature $kT$, normalization {\it norm} and 
metal abundance $Z$ of the ICM component.
In the outer regions ($>\timeform{25.2'}$), we fixed the ICM metal abundance to 0.2 as reported by \citet{kawano09}.

The 0.1--2.4 keV luminosity of A~3376
  ($L_X=2.5\times10^{44} \rm \ erg~s^{-1}$) is less than 1/3 of 
  the A~3667 value ($L_X=8.8\times10^{44} \rm \ erg~s^{-1}$), 
  based on the ROSAT X-ray-brightest Abell-type clusters survey~\citep{ebeling96}.  
  To examine the effect of flux
  contamination from different spatial regions, we estimated mutual
  photon contributions among all the annuli (see Appendix.\ 1).  The
  resultant photon fractions show that most of the detected photons
  come from the ``on-source'' region and about 20\% is from the adjacent
  annular regions.  These flux contamination features are very similar
  to that in A~3667~\citep{akamatsu11b}, in which the effect of stray
  light is always smaller than the systematic error of NXB\@.  

Recent study of the stray light properties of Suzaku
  XRT \citep{takei11} showed that its effect was significantly lower
  when the observing roll angle was inclined by 45 degrees just as in
  the case of the A~3376 west relic pointing.  Based on this estimation,
  we can expect that the effect of the stray light does not influence
  the result significantly.  We will not deal with the stray light any
  further in this paper.  
\begin{table*}[t]
\caption{ Best-fit background parameters}
\begin{center}
\begin{tabular}{lcccccccccccc}
\hline \hline
     			  & NOMINAL & CXBMAX &CXBMIN &CONTAMI+10\% &CONTAMI-10\% \\ \hline
 LHB& \\
 $kT (\rm keV)$ &      0.08 (fix)   &0.08 (fix)   &0.08 (fix)   &0.08 (fix)   &0.08 (fix)   &\\
$norm^{\ast}$  ($\times 10^{-3}$)&    $ {11.3}^{+3.4}_{-3.8} $  & $ {10.6}^{+3.2}_{-3.7} $	& $ {11.4}^{+3.2}_{-3.4} $ & $ {9.2}^{+3.3}_{-4.5} $&  $ {6.9}^{+3.2}_{-3.4} $ &\\ \hline
 MWH &                 \\
 $kT (\rm keV)$ &    $  {0.27}^{+0.06}_{-0.03}  $ &  $  {0.28}^{+0.06}_{-0.04}  $	&	 $  {0.28}^{+0.06}_{-0.04}  $&	  $  {0.27}^{+0.06}_{-0.04}  $	&  $  {0.27}^{+0.06}_{-0.04}  $  \\  
$norm^{\ast}$($\times 10^{-4}$)&   $  {4.1}^{+1.9}_{-1.5}  $  &  	$  {4.6}^{+1.9}_{-1.4}  $&$  {4.7}^{+1.9}_{-1.4}  $ &  $  {5.2}^{+2.2}_{-1.4}  $	& $  {4.0}^{+1.8}_{-1.4}  $  & \\ \hline
$\chi^{2}$/d.o.f    &  143 / 135& 147 / 135	& 152 / 135	& 145 / 135	&143 / 135&  \\ \hline
\multicolumn{6}{l}{\footnotesize
*:~Normalization of the apec component scaled with a factor 1/400$\pi$.}\\
\multicolumn{6}{l}{\footnotesize
Norm=$\rm \frac{1}{400\pi}$$\int n_{e}n_{H} dV/(4\pi(1+z^2)D_{A}^2)\times 10 ^{-14} \rm ~cm^{-5}~arcmin^{-2}$, where $D_A$ is the angular diameter distance to the source.}\\
\end{tabular}
\label{tab:bgd}
\end{center}
\end{table*}

\begin{table*}[t]
\begin{center}
\small
\caption{Estimation of the CXB fluctuation.}
\begin{tabular}{cccccccccccccc}\hline
 Region   & \timeform{0'}-\timeform{2.0'} & \timeform{2.0'}-\timeform{4.0'} &\timeform{4.0'}-\timeform{6.0'} &\timeform{6.0'}-\timeform{9.0'} & \timeform{12.0'}-\timeform{15.0'} & \timeform{15.0'}-\timeform{18'} & \timeform{18'}-\timeform{21'} & \timeform{21'}-\timeform{24'}  &   \timeform{24.'}-\timeform{27'}		&\timeform{27'}-\timeform{31'}		\\ \hline
 $SRR^{\ast}$    
&4.01
&3.35 
&3.29 
&3.40 
&0.36 
&0.76 
&1.08 
&0.95 
&0.83 
&0.66 \\
$\Omega_{\rm e,Suzaku}^{\dagger}$  
&12.4
&18.0
&30.3
&50.2
&12.6
&34.6
&60.7
&64.5
&67.0
&63.7 \\
${\sigma/I_{\rm CXB}}\ddagger$      
&47.1
&39.1 
&30.1 
&23.1 
&28.2 
&21.3 
&20.6 
&20.3 
&20.8
&20.2
\\
\hline
\multicolumn{6}{l}{$\ast:SOURCE\_RATIO\_REG$[\%] (See \cite{akamatsu11a}~Sec~3.4)}&\\
\multicolumn{1}{l}{$\dagger$:~[arcmin$^{2}$]} &\\
\multicolumn{5}{l}{$\ddagger:~S_{c}=5 \times10^{-14}$ erg cm$^{-2}$ s$^{-1}$ is assumed for all regions.}\\
\label{tab:cxb_fluc}
\end{tabular}
\end{center}
\end{table*}

\subsection{Estimation of the Background Components}\label{sec:bgd}
To exclude the point source contamination, we use {\it wavdetect} tool
in CIAO package version 4.3.0\footnote{http://cxc.harvard.edu/ciao/}
to identify point-like sources.   We adopt the value of the {\it sigthresh} parameter as $5\times10^{-6}$ in {\it wavdetect}, 
which is the significance threshold for the source detection.
As shown by magenta circles in Fig.\ \ref{fig:suzaku_image}, 
we detected 12 point-like sources and masked out the regions with 1 arcmin radius from the sources.  
The lowest detected 2--10 keV flux of these sources was $S_{c}=5 \times10^{-14}$ erg cm$^{-2}$ s$^{-1}$.  
Since this flux limit is lower than the level of $S_{c}=2 \times10^{-13}$ erg cm$^{-2}$ s$^{-1}$ by \citet{kushino02}, 
we adopted the CXB intensity as $5.97\times 10^{-8}~\rm \ erg ~cm^{-2}~s^{-1}~sr^{-1}$ after these point sources were subtracted.  
We also estimated the fluctuation of the CXB intensity based on ~\citet{akamatsu11a}.  
Table~\ref{tab:cxb_fluc} shows the resultant CXB fluctuation assuming the flux limit as $S_{c}=5 \times10^{-14}$ erg cm$^{-2}$ s$^{-1}$.

Since A~3376 is located far from the Galactic center ($l=246.53^\circ,
b=-26.29^\circ$), we did not consider contribution from the emission
associated with the Galactic plane.  To estimate the Galactic
background, we use another Suzaku observation data for Q~0551-3637
(Observation ID: 703036020, Hereafter Q0551) at $3^\circ.8$ from
A~3376 as shown in Table \ref{tab:obslog}.  
We excluded the region of the main source Q0551 with a radius of $3'$.
Using the data in the Q0551 field, we evaluated the model of the
Galactic background described in Sec~\ref{sec:spec_model} with $N_{\rm
  H}=3.2 \times 10^{20}$\ cm$^{-2}$.  For the fitting, we fixed the
temperature of the LHB component to 0.08 keV\@.  The spectrum was well
fitted with the above model ($\chi^{2}_\nu=1.08$ for 135 degrees of
freedom).  The resultant parameters are shown in Table~\ref{tab:bgd}.
The temperature of the MWH is $0.27^{+0.06}_{-0.03}$ keV, and the
temperature and intensity are consistent with the typical Galactic emission.  
For the estimation of the systematic error of the background spectrum, we adopted the CXB fluctuation as 15\%,
  and the OBF contamination to be $\pm 10\%$, respectively.  

In the next section, we also examined the effect of systematic errors
on our spectral parameters.  We considered the systematic error for
the NXB intensity to be $\pm 4.5~\%$~\citep{tawa08}, the fluctuation
of the CXB as shown in Table~\ref{tab:cxb_fluc}, and the OBF
contamination to be $\pm 10\%$, respectively.  The resultant
parameters after taking into account the systematic errors are shown
in Table~\ref{tab:bgd}.

\subsection{Single-Temperature Fit}

\begin{figure*}[htbp]
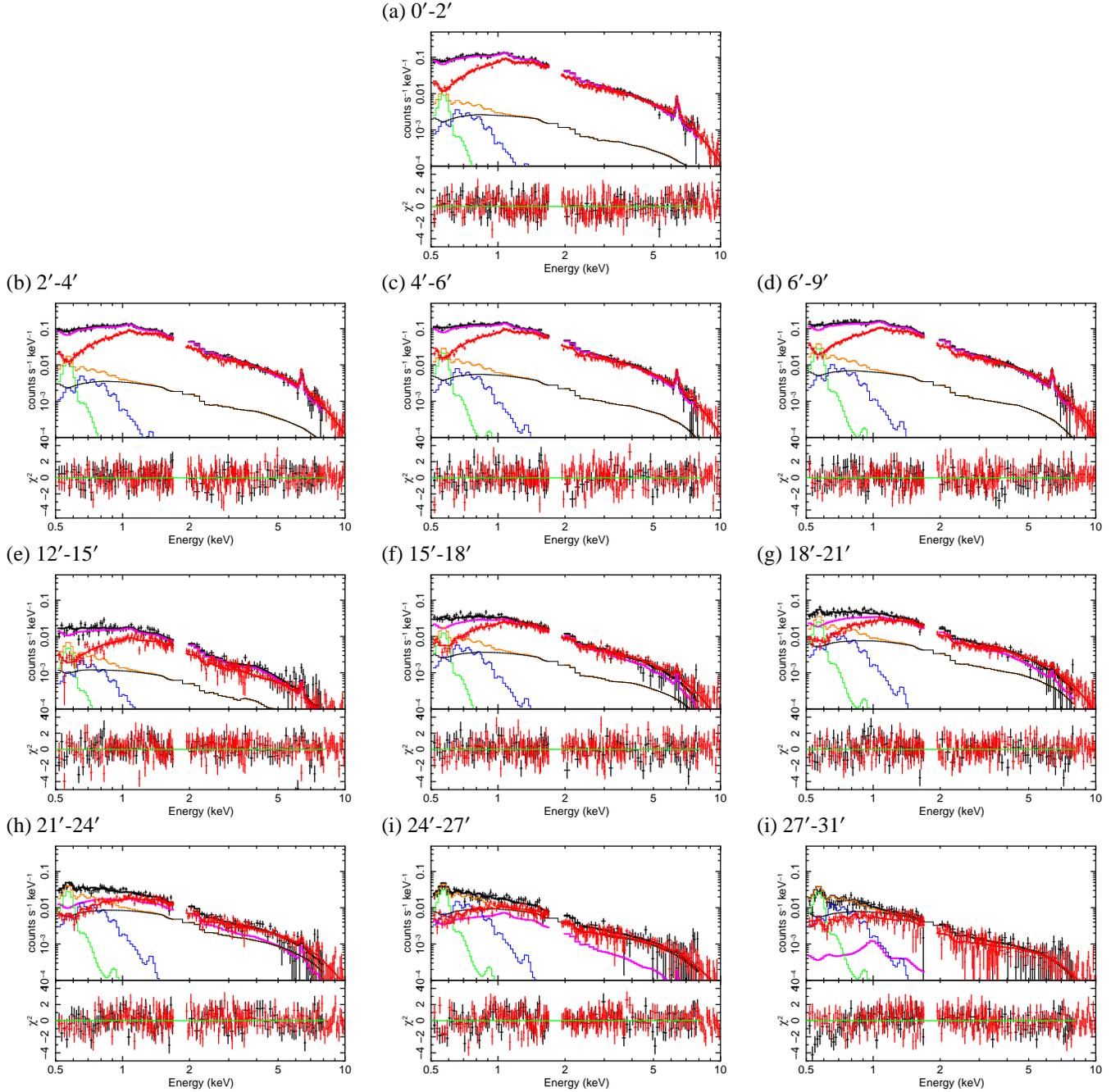

\begin{tabular}{cc}
\begin{minipage}{0.333\hsize}
(a)~\timeform{0'}-\timeform{2'}
\\[-0.8cm]
\begin{center}
 \includegraphics[angle=-90,scale=0.22]{./ICM-check-a3376e-1.ps}
\end{center}
\end{minipage}\\
\begin{minipage}{0.3333\hsize}
(b)~\timeform{2'}-\timeform{4'}
\\[-0.8cm]
\begin{center}
 \includegraphics[angle=-90,scale=0.22]{./ICM-check-a3376e-2.ps}
\end{center}
\end{minipage}
\begin{minipage}{0.3333\hsize}
(c)~\timeform{4'}-\timeform{6'}
\\[-0.8cm]
\begin{center}
 \includegraphics[angle=-90,scale=0.22]{./ICM-check-a3376e-3.ps}
\end{center}
\end{minipage}
\begin{minipage}{0.3333\hsize}
(d)~\timeform{6'}-\timeform{9'}
\\[-0.8cm]
\begin{center}
 \includegraphics[angle=-90,scale=0.22]{./ICM-check-a3376e-4.ps}
\end{center}
\end{minipage}\\
\begin{minipage}{0.3333\hsize}
(e)~\timeform{12'}-\timeform{15'}
\\[-0.8cm]
\begin{center}
 \includegraphics[angle=-90,scale=0.22]{./ICM-check-a3376a-5.ps}
\end{center}
\end{minipage}
\begin{minipage}{0.33333\hsize}
(f)~\timeform{15'}-\timeform{18'}
\\[-0.8cm]
\begin{center}
 \includegraphics[angle=-90,scale=0.22]{./ICM-check-a3376a-6.ps}
\end{center}
\end{minipage}
\begin{minipage}{0.3333\hsize}
(g)~\timeform{18'}-\timeform{21'}
\\[-0.8cm]
\begin{center}
 \includegraphics[angle=-90,scale=0.22]{./ICM-check-a3376a-7.ps}
\end{center}
\end{minipage}\\
\begin{minipage}{0.3333\hsize}
(h)~\timeform{21'}-\timeform{24'}
\\[-0.8cm]
\begin{center}
 \includegraphics[angle=-90,scale=0.22]{./ICM-check-a3376a-8.ps}
\end{center}
\end{minipage}
\begin{minipage}{0.33333\hsize}
(i)~\timeform{24'}-\timeform{27'}
\\[-0.8cm]
\begin{center}
 \includegraphics[angle=-90,scale=0.22]{./ICM-check-a3376a-9.ps}
\end{center}
\end{minipage}
\begin{minipage}{0.33333\hsize}
(i)~\timeform{27'}-\timeform{31'}
\\[-0.8cm]
\begin{center}
 \includegraphics[angle=-90,scale=0.22]{./ICM-check-a3376a-10.ps}
\end{center}
\end{minipage}
\end{tabular}
\caption{
NXB subtracted spectra in each annular region.
The XIS BI (black) and FI (red) spectra are fitted with the ICM model ({\it wabs + apec}),
along with the sum of the CXB and the Galactic emission ({\it apec + wabs(apec + powerlaw)}).
The CXB component is shown with a black curve, and the LHB and MWH emissions are indicated
by green and blue curves, respectively.
The total background components are shown by the orange curve.
}
\label{fig:fit}
\end{figure*}

The pulse-height spectra and the best-fit models for all the annular
regions are shown in Fig.\ \ref{fig:fit}.  The parameters and the
resultant $\chi^2$ values are listed in Table \ref{tab:fit}.
Considering both the statistical and systematic errors, 
the ICM signal was significantly detected out to $31'$ (1.67 Mpc $\sim 0.9 r_{200}$) for the first time.

\begin{figure*}[]
\begin{tabular}{cc}
\begin{minipage}{0.5\hsize}
(a) Temperature
\\[-0.8cm]
\begin{center}
\includegraphics[scale=0.35,angle=-90]{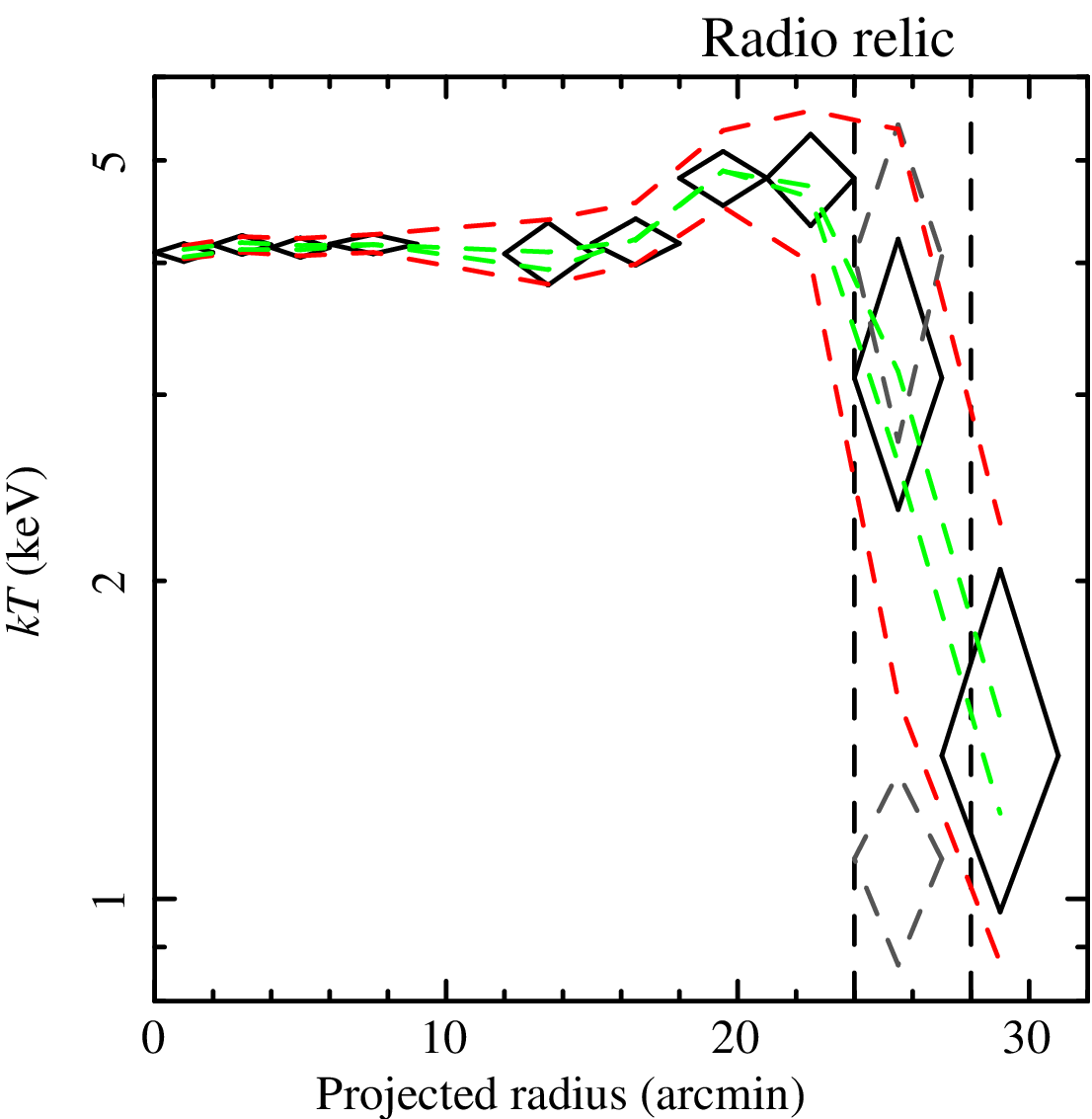}
\end{center}
\end{minipage}
\begin{minipage}{0.5\hsize}
(b) Surface brightness
\\[-0.8cm]
\begin{center}
\includegraphics[scale=0.35,angle=-90]{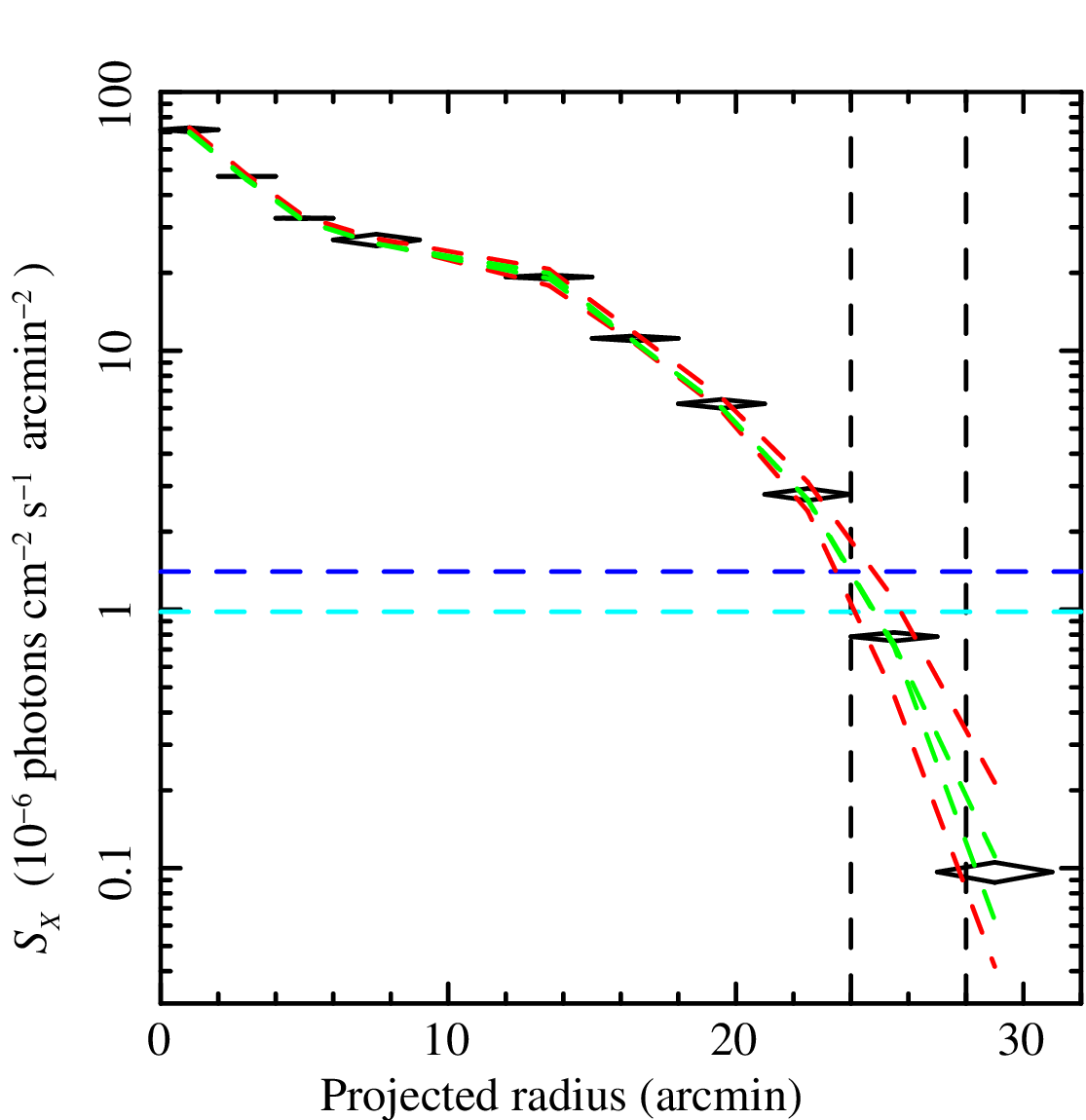}
\end{center}
\end{minipage}\\
\begin{minipage}{0.5\hsize}
(c) Deprojected electron density
\begin{center}
\includegraphics[scale=0.35,angle=-90]{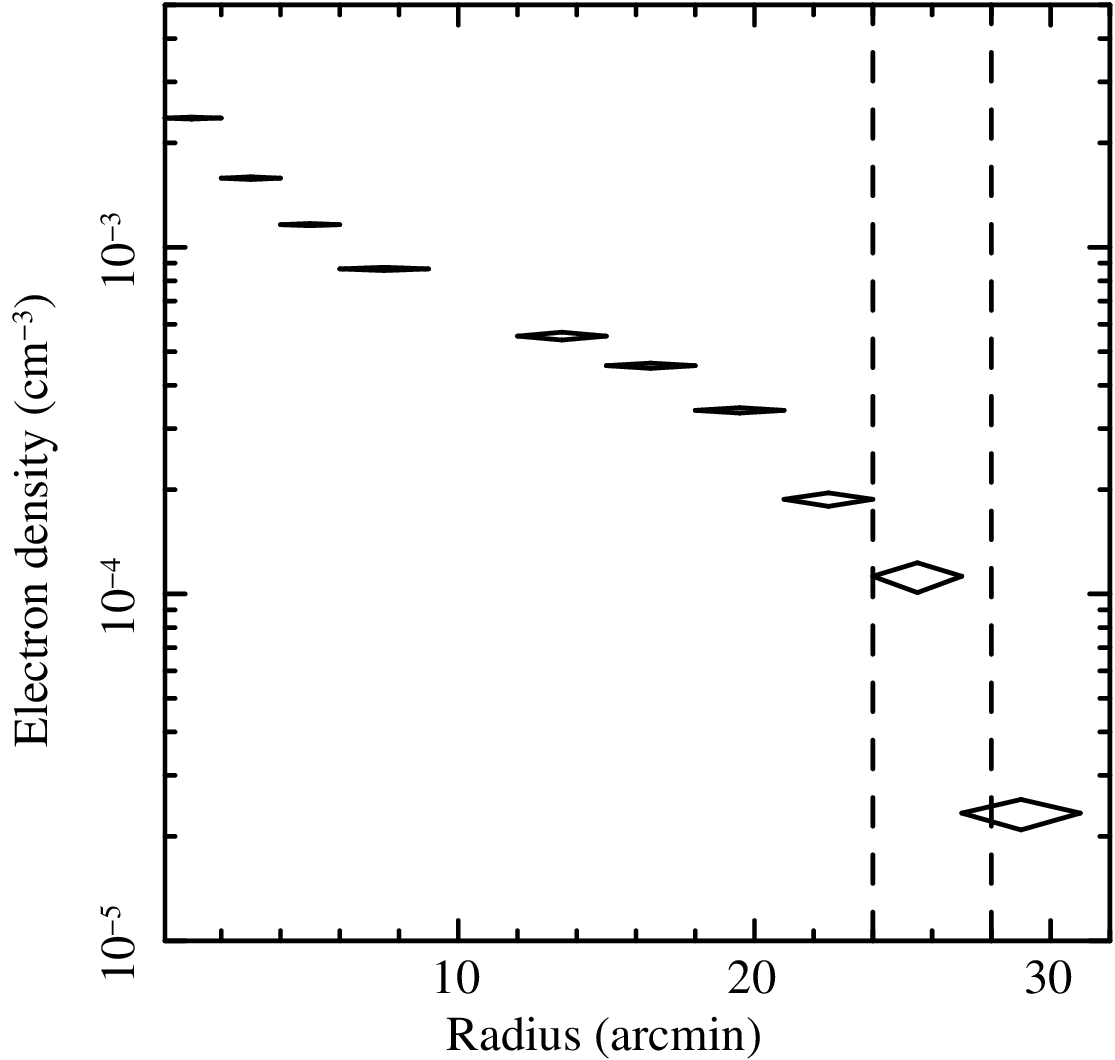}
\end{center}
\end{minipage}
\begin{minipage}{0.5\hsize}
(d) Pressure
\begin{center}
\includegraphics[scale=0.35,angle=-90]{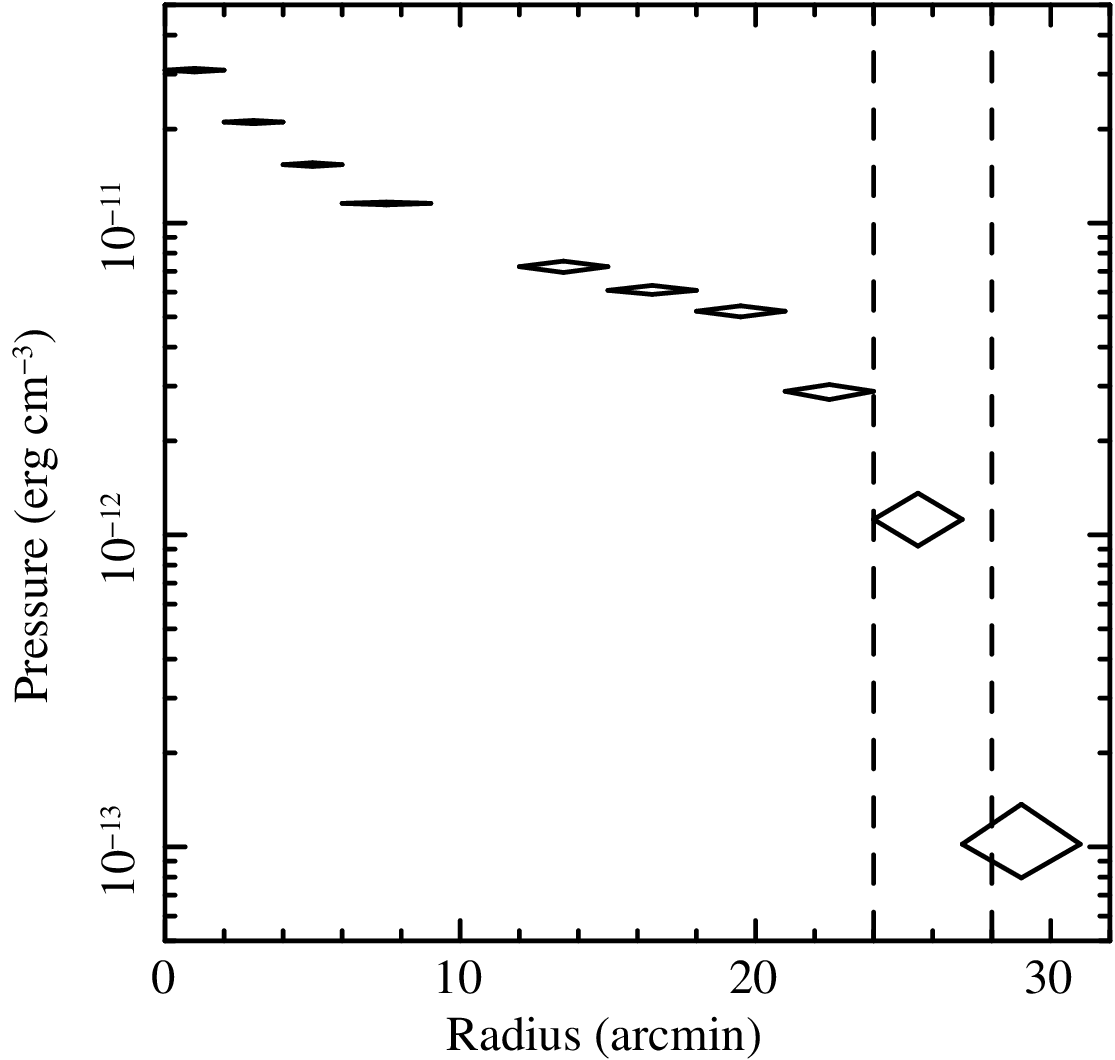}
\end{center}
\end{minipage}
\end{tabular}
\caption{ 
Radial profiles of (a) ICM temperature, (b) surface brightness, 
(c) deprojected electron density, and (d) gas pressure. Suzaku best fit values 
with statical errors
are shown with black diamonds.  
 Black dashed vertical lines show approximate radial boundaries of the west radio relic. 
 Green and red dashed lines show typical systematic changes of the best-fit values due to change of OBF contaminants 
 and NXB level.  
 In temperature profile, gray dashed diamonds show the result of $2kT$ model fitting (see text).  
 In surface brightness profile, blue and cyan dashed lines show the Galactic background and CXB emission, respectively.  
 }
\label{fig:profile}
\end{figure*}

Fig.\ \ref{fig:profile} shows radial profiles of ICM parameters
(temperature, surface brightness, deprojected electron density, and
thermal pressure).  The temperature within $20'$ (1.2 Mpc corresponding to $0.6r_{200}$) of the cluster center shows an
  almost flat or rising profile with $kT \sim 4.0$ keV, which is very
  much different from those seen in the relaxed clusters.  
The increase of temperature just inside of the radio relic
(\timeform{18'}-\timeform{21'}, and \timeform{21'}-\timeform{24'})
compared with the level in the inner region is significant by
about $4 \sigma$.  The temperature of A~3376 suddenly drops at the radio
relic region from 4.7 keV to 1.3 keV\@.  These features strongly
suggest that there is a shock heating taking place in the relic
region.  The surface brightness profile also shows more than an order
of magnitude drop at the same region.

Based on the same Suzaku data, \citet{kawano09} showed the temperature in the radio relic region to be $kT \approx 3.8$ keV\@.  
The region roughly corresponds to a combination of $21'-24'$ and $24'-27'$ in our analysis, and our temperatures of 
4.7 keV and 3.1 keV for the two regions indicate that our results are consistent with the previous study.  

We also calculated the thermal pressure profile, with the pressure
defined as $P=n_{e}kT$\@.  The deprojected electron density
$n_e$ was derived from the {\it apec} normalization as in
\citet{akamatsu11a}.  As seen in Fig.\ \ref{fig:profile}, the
deprojected electron density and pressure profiles show a remarkable
drop across the radio relic, which strongly suggests existence of the
shock front.

\begin{table*}[ht]
\begin{center}
\footnotesize
\caption{Best-fit parameters of the ICM}
\begin{tabular}{cccccccccccc}\hline
 & \timeform{0'}-\timeform{2'} &\timeform{2'}-\timeform{4'}   &$\timeform{4'}-\timeform{6'}$ &$\timeform{6'}-\timeform{9'}$ & $\timeform{12'}-\timeform{15'}$ & $\timeform{15'}-\timeform{18'}$ & $\timeform{18'}-\timeform{21'}$ & $\timeform{21'}-\timeform{24'}$ & $\timeform{24'}-\timeform{27'}$ & $\timeform{27'}-\timeform{31'}$\\ \hline
 \multicolumn{11}{c}{NOMIMAL}\\ \hline
$kT$ (keV)&
 $  {4.09}^{+0.08}_{-0.08}  $  & 
 $  {4.16}^{+0.09}_{-0.09}  $  & 
 $  {4.14}^{+0.09}_{-0.09}  $  & 
 $  {4.17}^{+0.09}_{-0.09}  $  & 
 $  {4.08}^{+0.29}_{-0.27}  $  & 
 $  {4.17}^{+0.23}_{-0.19}  $  & 
 $  {4.81}^{+0.29}_{-0.28}  $  & 
 $  {4.68}^{+0.48}_{-0.48}  $  & 
 $  {3.11}^{+1.10}_{-0.78}  $  & 
 $  {1.34}^{+0.69}_{-0.39}  $  & 

 \\
 $Z$&
 $  {0.33}^{+0.03}_{-0.03}  $  & 
 $  {0.28}^{+0.03}_{-0.03}  $  & 
 $  {0.26}^{+0.03}_{-0.03}  $  & 
 $  {0.25}^{+0.03}_{-0.03}  $  & 
 $  {0.21}^{+0.09}_{-0.08}  $  & 
 $  {0.18}^{+0.06}_{-0.06}  $  & 
 $  {0.11}^{+0.06}_{-0.06}  $  & 
 0.2 (fix)& 
 0.2 (fix) & 
 0.2 (fix)  & 
 \\
norm$^{\ast}$&
 $  {164.8}^{+3.2}_{-3.2}  $  & 
 $  {108.3}^{+1.9}_{-1.9}  $  & 
 $  {75.0}^{+1.1}_{-1.1}  $  & 
 $  {64.7}^{+1.3}_{-1.3}  $  & 
 $  {46.4}^{+2.3}_{-2.3}  $  & 
 $  {25.5}^{+0.9}_{-0.9}  $  & 
 $  {15.3}^{+0.5}_{-0.5}  $  & 
 $  {6.7}^{+0.3}_{-0.3}  $  & 
 $  {2.1}^{+0.2}_{-0.2}  $  & 
 $  {0.3}^{+0.2}_{-0.2}  $  & 
  \\
 $S^{\dagger}_{\rm 0.4-10\ keV}$&
 $  {70.11}^{+1.45}_{-1.45}  $  & 
 $  {46.28}^{+0.13}_{-0.13}  $  & 
 $  {31.88}^{+0.14}_{-0.14}  $  & 
 $  {26.25}^{+1.41}_{-1.41}  $  & 
 $  {19.04}^{+0.48}_{-0.48}  $  & 
 $  {10.92}^{+0.28}_{-0.28}  $  & 
 $  {6.14}^{+0.22}_{-0.22}  $  & 
 $  {2.72}^{+0.15}_{-0.15}  $  & 
 $  {0.77}^{+0.04}_{-0.04}  $  & 
 $  {0.13}^{+0.01}_{-0.01}  $  & 
 \\
 $\chi^{2}$/d.o.f &
 322 / 307  & 
 314 / 307  & 
 388 / 307  & 
 342 / 307  & 
 362 / 307  & 
 352 / 307  & 
 348 / 307  & 
 351 / 308  & 
 345 / 308  & 
 361 / 308  & 

 \\ \hline
 \multicolumn{11}{c}{CXB MAX+NXB 4.5\% RED}\\ \hline
 $kT$ (keV)&
 $  {4.15}^{+0.08}_{-0.08}  $  & 
 $  {4.23}^{+0.09}_{-0.09}  $  & 
 $  {4.22}^{+0.09}_{-0.09}  $  & 
 $  {4.26}^{+0.11}_{-0.09}  $  & 
 $  {4.39}^{+0.35}_{-0.28}  $  & 
 $  {4.56}^{+0.25}_{-0.25}  $  & 
 $  {5.34}^{+0.36}_{-0.28}  $  & 
 $  {5.57}^{+0.59}_{-0.48}  $  & 
 $  {5.35}^{+1.45}_{-1.00}  $  & 
 $  {2.27}^{+1.65}_{-1.91}  $  & 
   \\
  $Z$&
 $  {0.33}^{+0.03}_{-0.03}  $  & 
 $  {0.28}^{+0.03}_{-0.03}  $  & 
 $  {0.26}^{+0.03}_{-0.03}  $  & 
 $  {0.26}^{+0.03}_{-0.03}  $  & 
 $  {0.20}^{+0.09}_{-0.08}  $  & 
 $  {0.19}^{+0.06}_{-0.06}  $  & 
 $  {0.12}^{+0.06}_{-0.06}  $  & 
 0.2 (fix)& 
 0.2 (fix) & 
 0.2 (fix)  & 

  \\
norm$^{\ast}$&
 $  {164.8}^{+3.2}_{-3.2}  $  & 
 $  {108.3}^{+1.9}_{-1.9}  $  & 
 $  {76.1}^{+1.1}_{-1.1}  $  & 
 $  {65.4}^{+1.3}_{-1.3}  $  & 
 $  {49.3}^{+2.3}_{-2.3}  $  & 
 $  {26.2}^{+0.9}_{-0.9}  $  & 
 $  {15.8}^{+0.5}_{-0.5}  $  & 
 $  {7.3}^{+0.3}_{-0.3}  $  & 
 $  {2.6}^{+0.2}_{-0.2}  $  & 
 $  {0.5}^{+0.2}_{-0.2}  $  & 
  \\
 $S^{\dagger}_{\rm 0.4-10\ keV}$&
 $  {70.95}^{+1.48}_{-1.48}  $  & 
 $  {46.89}^{+0.13}_{-0.13}  $  & 
 $  {32.32}^{+0.14}_{-0.14}  $  & 
 $  {26.57}^{+1.41}_{-1.41}  $  & 
 $  {20.35}^{+0.55}_{-0.55}  $  & 
 $  {11.36}^{+0.28}_{-0.28}  $  & 
 $  {6.45}^{+0.24}_{-0.24}  $  & 
 $  {3.02}^{+0.17}_{-0.17}  $  & 
 $  {1.07}^{+0.06}_{-0.06}  $  & 
 $  {0.21}^{+0.02}_{-0.02}  $  & 

  \\
 $\chi^{2}$/d.o.f & 
 327 / 307  & 
 317 / 307  & 
 391 / 307  & 
 347 / 307  & 
 362 / 307  & 
 353 / 307  & 
 350 / 307  & 
 356 / 308  & 
 352 / 308  & 
 373 / 308  &

 \\ \hline
 \multicolumn{11}{c}{CXB MIN+NXB 4.5\% ADD}\\ \hline
  $kT$ (keV)&
 $  {4.04}^{+0.08}_{-0.08}  $  & 
 $  {4.09}^{+0.09}_{-0.09}  $  & 
 $  {4.06}^{+0.09}_{-0.09}  $  & 
 $  {4.09}^{+0.09}_{-0.09}  $  & 
 $  {3.82}^{+0.29}_{-0.29}  $  & 
 $  {3.99}^{+0.21}_{-0.20}  $  & 
 $  {4.52}^{+0.31}_{-0.28}  $  & 
 $  {4.00}^{+0.43}_{-0.40}  $  & 
 $  {1.55}^{+0.26}_{-0.21}  $  & 
 $  {0.87}^{+0.57}_{-0.47}  $  &
   \\
  $Z$&
 $  {0.33}^{+0.03}_{-0.03}  $  & 
 $  {0.28}^{+0.03}_{-0.03}  $  & 
 $  {0.26}^{+0.03}_{-0.03}  $  & 
 $  {0.25}^{+0.03}_{-0.03}  $  & 
 $  {0.22}^{+0.09}_{-0.09}  $  & 
 $  {0.18}^{+0.06}_{-0.06}  $  & 
 $  {0.12}^{+0.06}_{-0.06}  $  & 
  0.2 (fix)& 
 0.2 (fix) & 
 0.2 (fix)  & 
  \\
norm$^{\ast}$&
$  {161.6}^{+3.2}_{-3.2}  $  & 
 $  {106.4}^{+1.9}_{-1.9}  $  & 
 $  {72.8}^{+1.1}_{-1.1}  $  & 
 $  {63.4}^{+1.3}_{-1.3}  $  & 
 $  {42.7}^{+2.3}_{-2.3}  $  & 
 $  {24.2}^{+0.9}_{-0.9}  $  & 
 $  {14.0}^{+0.5}_{-0.5}  $  & 
 $  {5.7}^{+0.3}_{-0.3}  $  & 
 $  {1.1}^{+0.2}_{-0.2}  $  & 
 $  {0.1}^{+0.2}_{-0.1}  $  & 

  \\
 $S^{\dagger}_{\rm 0.4-10\ keV}$&
 $  {69.04}^{+1.37}_{-1.37}  $  & 
 $  {45.49}^{+0.11}_{-0.11}  $  & 
 $  {31.25}^{+0.11}_{-0.11}  $  & 
 $  {25.76}^{+1.37}_{-1.37}  $  & 
 $  {17.53}^{+0.39}_{-0.39}  $  & 
 $  {10.31}^{+0.28}_{-0.28}  $  & 
 $  {5.68}^{+0.19}_{-0.19}  $  & 
 $  {2.29}^{+0.12}_{-0.12}  $  & 
 $  {0.41}^{+0.02}_{-0.02}  $  & 
 $  {0.04}^{+0.01}_{-0.01}  $  & 

  \\
 $\chi^{2}$/d.o.f & 
 316 / 307  & 
 314 / 307  & 
 391 / 307  & 
 339 / 307  & 
 367 / 307  & 
 350 / 307  & 
 357 / 307  & 
 367 / 308  & 
 351 / 308  & 
 414 / 308  & 

 \\ \hline
  \multicolumn{11}{c}{CONTAMI 10\% ADD}\\ \hline
   $kT$ (keV)&
  $  {4.05}^{+0.08}_{-0.08}  $  & 
 $  {4.12}^{+0.09}_{-0.09}  $  & 
 $  {4.12}^{+0.09}_{-0.09}  $  & 
 $  {4.16}^{+0.09}_{-0.09}  $  & 
 $  {3.96}^{+0.27}_{-0.27}  $  & 
 $  {4.20}^{+0.25}_{-0.20}  $  & 
 $  {4.88}^{+0.29}_{-0.28}  $  & 
 $  {4.73}^{+0.48}_{-0.46}  $  & 
 $  {3.16}^{+1.19}_{-0.80}  $  & 
 $  {1.37}^{+1.14}_{-0.40}  $  & 

   \\
  $Z$&
 $  {0.32}^{+0.03}_{-0.03}  $  & 
 $  {0.28}^{+0.03}_{-0.03}  $  & 
 $  {0.26}^{+0.03}_{-0.03}  $  & 
 $  {0.26}^{+0.03}_{-0.03}  $  & 
 $  {0.21}^{+0.09}_{-0.08}  $  & 
 $  {0.18}^{+0.06}_{-0.06}  $  & 
 $  {0.12}^{+0.06}_{-0.06}  $  & 
  0.2 (fix)& 
 0.2 (fix) & 
 0.2 (fix)  & 
  \\
norm$^{\ast}$&
 $  {164.8}^{+3.2}_{-3.2}  $  & 
 $  {108.3}^{+1.9}_{-1.9}  $  & 
 $  {75.0}^{+1.1}_{-1.1}  $  & 
 $  {64.1}^{+1.3}_{-1.3}  $  & 
 $  {45.9}^{+2.3}_{-2.3}  $  & 
 $  {25.1}^{+0.9}_{-0.9}  $  & 
 $  {14.9}^{+0.5}_{-0.5}  $  & 
 $  {6.6}^{+0.3}_{-0.3}  $  & 
 $  {1.9}^{+0.2}_{-0.2}  $  & 
 $  {0.2}^{+0.1}_{-0.1}  $  & 
  \\
 $S^{\dagger}_{\rm 0.4-10\ keV}$&
 $  {70.12}^{+1.69}_{-1.69}  $  & 
 $  {46.10}^{+0.42}_{-0.42}  $  & 
 $  {31.69}^{+0.29}_{-0.29}  $  & 
 $  {25.92}^{+1.63}_{-1.63}  $  & 
 $  {34.94}^{+1.59}_{-1.59}  $  & 
 $  {10.77}^{+0.21}_{-0.21}  $  & 
 $  {6.03}^{+0.25}_{-0.25}  $  & 
 $  {2.64}^{+0.16}_{-0.16}  $  & 
 $  {0.72}^{+0.04}_{-0.04}  $  & 
 $  {0.06}^{+0.01}_{-0.01}  $  &
  \\
 $\chi^{2}$/d.o.f & 
 323 / 307  & 
 316 / 307  & 
 388 / 307  & 
 338 / 307  & 
 364 / 307  & 
 348 / 307  & 
 354 / 307  & 
 363 / 308  & 
 360 / 308  & 
 402 / 308  & 
 
 \\ \hline
    \multicolumn{11}{c}{CONTAMI 10\% RED}\\ \hline
     $kT$ (keV)&

  $  {4.12}^{+0.08}_{-0.08}  $  & 
 $  {4.18}^{+0.09}_{-0.09}  $  & 
 $  {4.15}^{+0.09}_{-0.09}  $  & 
 $  {4.16}^{+0.09}_{-0.09}  $  & 
 $  {3.94}^{+0.27}_{-0.27}  $  & 
 $  {4.21}^{+0.26}_{-0.20}  $  & 
 $  {4.89}^{+0.29}_{-0.28}  $  & 
 $  {4.61}^{+0.47}_{-0.43}  $  & 
 $  {2.59}^{+1.00}_{-0.58}  $  & 
 $  {1.29}^{+0.64}_{-0.31}  $  & 
   \\
  $Z$&
  $  {0.33}^{+0.03}_{-0.03}  $  & 
 $  {0.28}^{+0.03}_{-0.03}  $  & 
 $  {0.26}^{+0.03}_{-0.03}  $  & 
 $  {0.25}^{+0.03}_{-0.03}  $  & 
 $  {0.21}^{+0.09}_{-0.08}  $  & 
 $  {0.17}^{+0.06}_{-0.06}  $  & 
 $  {0.11}^{+0.06}_{-0.06}  $  & 
   0.2 (fix)& 
 0.2 (fix) & 
 0.2 (fix)  & 
  \\
norm$^{\ast}$&
 $  {161.6}^{+3.2}_{-3.2}  $  & 
 $  {106.4}^{+1.9}_{-1.9}  $  & 
 $  {73.9}^{+1.1}_{-1.1}  $  & 
 $  {64.7}^{+1.3}_{-1.3}  $  & 
 $  {46.1}^{+2.3}_{-2.3}  $  & 
 $  {25.3}^{+0.9}_{-0.9}  $  & 
 $  {15.1}^{+0.5}_{-0.5}  $  & 
 $  {6.6}^{+0.3}_{-0.3}  $  & 
 $  {2.0}^{+0.2}_{-0.2}  $  & 
 $  {0.3}^{+0.2}_{-0.2}  $  & 
  \\
 $S^{\dagger}_{\rm 0.4-10\ keV}$&
  $  {69.34}^{+1.75}_{-1.75}  $  & 
 $  {45.65}^{+0.48}_{-0.48}  $  & 
 $  {31.53}^{+0.34}_{-0.34}  $  & 
 $  {25.92}^{+1.66}_{-1.66}  $  & 
 $  {35.00}^{+1.59}_{-1.59}  $  & 
 $  {10.76}^{+0.19}_{-0.19}  $  & 
 $  {6.03}^{+0.26}_{-0.26}  $  & 
 $  {2.66}^{+0.16}_{-0.16}  $  & 
 $  {0.73}^{+0.04}_{-0.04}  $  & 
 $  {0.11}^{+0.01}_{-0.01}  $  & 

  \\
 $\chi^{2}$/d.o.f &
  317 / 307  & 
 315 / 307  & 
 390 / 307  & 
 338 / 307  & 
 362 / 307  & 
 347 / 307  & 
 349 / 307  & 
 349 / 308  & 
 333 / 308  & 
 356 / 308  & 

  \\ \hline
 \multicolumn{10}{l}{
*:~Normalization of the {\it apec} component scaled with a factor
SOURCE-RATIO-REG/$\Omega_e$ from Table~\ref{tab:cxb_fluc},
}\\
\multicolumn{10}{l}{
Norm=$\frac{\rm SOURCE-RATIO-REG}{\Omega_e}\int n_{e}n_{H}
dV/(4\pi(1+z^2)D_{A}^2)\times 10
^{-20} ~{\rm cm}^{-5}~{\rm arcmin}^{-2}$,
where $D_A$ is the angular diameter distance to the source.}\\
\multicolumn{10}{l}{
$\dagger$: Photon flux in units of $10^{-6} \rm ~photons~ cm^{-2}~ s^{-1}~ arcmin^{-2}$. Energy band is 0.4 - 10.0 keV\@.
}\\
\multicolumn{10}{l}{
Surface brightness of the apec component scaled with a factor
SOURCE-RATIO-REG and $\Omega_e$ from Table~\ref{tab:cxb_fluc}.
}
\label{tab:fit}
\end{tabular}
\end{center}
\end{table*}

\begin{table}[t]
\begin{center}
\caption{ The best fit parameters of 2 $kT$ model with
statistical and systematic errors (CXB fluctuations and OBF contamination)} \label{tab:2kt}
\begin{tabular}{cccccccc}\hline
&				2 $kT$			\\ \hline
$kT_{\rm high}$ 	&	4.05$\pm1.62\pm1.37\pm0.11$	\\
$kT_{\rm low}$ 	&	1.09$\pm0.22\pm0.25\pm0.07$	\\
$Z$			&	0.2 (fix)				\\
$norm_{\rm high}^\ast$&1.$56\pm0.6\pm0.7\pm0.3$						\\
$norm_{\rm low}^\ast$	&0.12$\pm0.03\pm0.8\pm0.2$			\\
$\chi^2$/ d.o.f&	329/306				\\ \hline
\multicolumn{3}{l}{\footnotesize $\ast$: Same as Table~\ref{tab:fit}.} \\
\multicolumn{3}{l}{\footnotesize $\dagger$: 10-40 keV  power-law flux in units of 
} \\
\multicolumn{3}{l}{\footnotesize  10$^{-14}$~\fluxunit \rm ~arcmin$^{-2}$}
\end{tabular}
\end{center}
\end{table}

\subsection{Spectral Features in the Radio Relic Region}\label{sec:multi}
The radio relic region was mostly covered by the deep Suzaku
observation ($\sim 100$ ks).  In the radio relic of A~3667,
\citet{akamatsu11b} examined the possibility of multi-temperature
plasma using $2 kT$ model and several $DEM$ (Differential Emission
Measure: \cite{kaastra96}) models.  They confirmed that the emission
measure distribution is characterized by two peaks around the radio
relic region.  The sharp drop of temperature in the radio relic of
A~3376 suggests that a multi-temperature model is needed to describe
the spectrum. The data will allow us to examine the relative
contributions from different temperature components.

To examine the multi-temperature feature in the radio relic region
(\timeform{24'}-\timeform{27'}), we included additional thermal
component ($2 kT$) to the single temperature model.  We then obtained
an almost acceptable fit with $\chi^2$/ d.o.f.\ = 329/306
(Table~\ref{tab:2kt}) compared with the single temperature case of
345/308. 
The rather low statics of the spectrum hampered us to distinguish between the $2 kT$
and $1 kT$ models.  The hot component temperature was derived as
$kT_{\rm high}=4.05\pm 1.62$~keV, and the cool one as $kT_{\rm
  low}=1.09\pm 0.22$~keV, while the common abundance was fixed to 0.2
solar.  These values seem reasonable considering the result of the
single temperature fit, which gave $kT=3.11\pm{1.10}$~keV\@.  These
two temperature values are close to those observed inside and outside
of the radio relic regions.

The actual three dimensional structure of the shock front would be a
``bow'' shape.  Considering that the observed data should contain both
the upstream and downstream components with respect to the shock front,
it is natural to expect that the data in the radio relic region of
A~3376 is a mixture of high and low temperature ICM by the projection
effect. The ratio of emission measures can be used to constrain the
shock front structure, however we leave such a study for future works.

\begin{table*}[]
\begin{center}
\caption{ The best fit parameters of 1$kT$ +PL model with
statistical error} \label{tab:relic}
\begin{tabular}{cccccccc}\hline
	&\multicolumn{2}{c}{ $\timeform{18'}-\timeform{21'}$ }	&\multicolumn{2}{c}{ $\timeform{21'}-\timeform{24'}$ }	&\multicolumn{2}{c}{ $\timeform{24'}-\timeform{27'}$ }\\ 
&				PL~$(\Gamma=2.0)$		&PL~$(\Gamma=2.5)$ &	PL~$(\Gamma=2.0)$		&PL~$(\Gamma=2.5)$ &PL~$(\Gamma=2.0)$		&PL~$(\Gamma=2.5)$ \\ \hline
$kT$ 			& $4.94_{-0.28}^{+0.29}$		&4.96$_{-0.29}^{+0.32}$&4.80$_{-0.46}^{+0.47}$	&4.78$_{-0.43}^{+0.49}$	&	3.11$_{-0.67}^{+1.26}$		&3.10$_{-0.64}^{+1.05}$\\
$Z$				&0.12$_{-0.06}^{+0.06}$		&0.117$_{-0.06}^{+0.06}$&0.26$_{-0.11}^{+0.13}$	&0.27$_{-0.12}^{+0.12}$						&	0.2 (fix)					&	0.2 (fix)		\\
$norm^\ast$		&14.5$_{-0.4}^{+0.5}$		&	14.6$_{-0.6}^{+0.6}$	&6.2 $_{-0.3}^{+0.5}$	&	6.3	 $_{-0.3}^{+0.3}$		&	 1.9$_{-0.3}^{+0.2}$		&	1.9$_{-0.2}^{+0.3}$		\\
$\Gamma$		&	 2.0 (fix)					&	2.5 (fix)	&	 2.0 (fix)					&	2.5 (fix)	&	 2.0 (fix)					&	2.5 (fix)		\\
Flux UL$^{\dagger}$&		1.02	&    1.06	&		0.99		&  1.03				&	0.99					&	1.08		\\
$\chi^2$/ d.o.f		&351/306		&	351/306			&356/306	&355/306				&	347/307					&	348/307	\\ \hline
\multicolumn{3}{l}{\footnotesize $\ast$: Same as Table~\ref{tab:fit}.} \\
\multicolumn{6}{l}{\footnotesize $\dagger$: 10-40 keV flux of the power-law component, \footnotesize in 10$^{-14}$~\fluxunit \rm arcmin$^{-2}$}
\end{tabular}
\end{center}
\end{table*}

\subsection{Constraint on Non-Thermal Emission}\label{sec:IC}

We evaluated the flux of the possible non-thermal X-ray emission using
three XIS spectra around the radio relic region (Fig.\
\ref{fig:fit}g, h, i).  \citet{kawano09} gave the $3\sigma$ upper
limit flux for a power-law component in the radio relic region to be
$1.1\times 10^{-12}$ erg cm$^{-2}$ s$^{-1}$ in 4--8 keV and $1.9\times
10^{-12}$ erg cm$^{-2}$ s$^{-1}$ with extrapolation to 15--50 keV,
respectively, assuming the photon index to be 2.0.  We added a
power-law component to the above ICM model as: {\it apec$_{\rm LHB}$ +
  wabs {\rm(}apec$_{\rm MWH}$ + powerlaw$_{\rm CXB}$ + apec$_{\rm
    ICM}$ + powerlaw$_{\rm Non-thermal}${\rm )}.}  The dependence on
the photon index was examined by adopting 2.0 and 2.5 in the fit.

The results of the spectral fits are summarized in
Table~\ref{tab:relic}. We found no significant difference against the
photon index, and all three regions showed only upper
  limits for the power-law component. 
The derived upper-limit fluxes are $F_{\rm 10-40\ keV}<1.1\times10^{-14}$ \fluxunit\ arcmin$^{-2}$  at $\Gamma=2.5$. 
To compare with the result by \citet{kawano09}, we also derived the 4-8 keV flux as
$F_{\rm 4-8\ keV}<5.2\times10^{-15}$ \fluxunit\ arcmin$^{-2}$
Using the solid angle covered by the radio relic
($\Omega_{\rm relic}=122$ arcmin$^2$), the flux upper limit in the
4-8 keV band is $5.3\times 10^{-13}$ \fluxunit, which is consistent with the previous Suzaku results
with XIS \& HXD \citep{kawano09}. 
We note that the radial temperature profile shows an increase just inside of the 
radio relic over the level of the central region by about 1  keV\@. 
The result of the spectral fit implies that this increase is not due to the presence of a hard power-law emission 
but rather showing the pure ICM heating caused by the shock.

\section{Discussion}
\label{sec:discussion}
Suzaku performed 2 pointing observations of Abell 3376 along its
merger axis.  The ICM temperature was found to be fairly flat at about
4 keV with a slow rise with radius in the region from the center to
$0.6 r_{200}$~(1 Mpc).  This is followed by a sharp drop in the radio
relic region, which implies there is a shock front.  We will attempt
to evaluate the gas properties related with the shock (temperature,
Mach number, magnetic field, and entropy) and discuss their
implications.
\subsection{Temperature Profiles}\label{sec:kt_comp}
The ICM temperature is an important observational
  parameter which reflects the depth of the gravitational potential of
  the cluster.  Their profiles also enable us to look into the history
  of the ICM heating and the growth of clusters.  

Previous Chandra, XMM-Newton results for other clusters indicated a
significant temperature decline toward the outer regions.  Recent
studies with Suzaku showed temperature profiles of ICM to the virial
radius ($r_{200}$) for several clusters \citep{george08, reiprich09,
  bautz09, hoshino10, kawaharada10, simionescu11}.  These profiles for
relaxed clusters are approximated by a ``universal'' profile which
declines by a factor of $\sim 3$ at $r_{200}$ from the center.  On the
other hand, merging clusters show complex temperature distribution
depending on the mass ratio and collision geometry.  In the outer
regions toward $r_{200}$, radial temperature profiles show some
excesses and sharp drops which are not seen in relaxed systems
(\cite{akamatsu11a,akamatsu11b}).  
However, very little systematic studies have been made about 
the ICM properties in the outer regions of merging clusters.

\begin{figure*}[t]
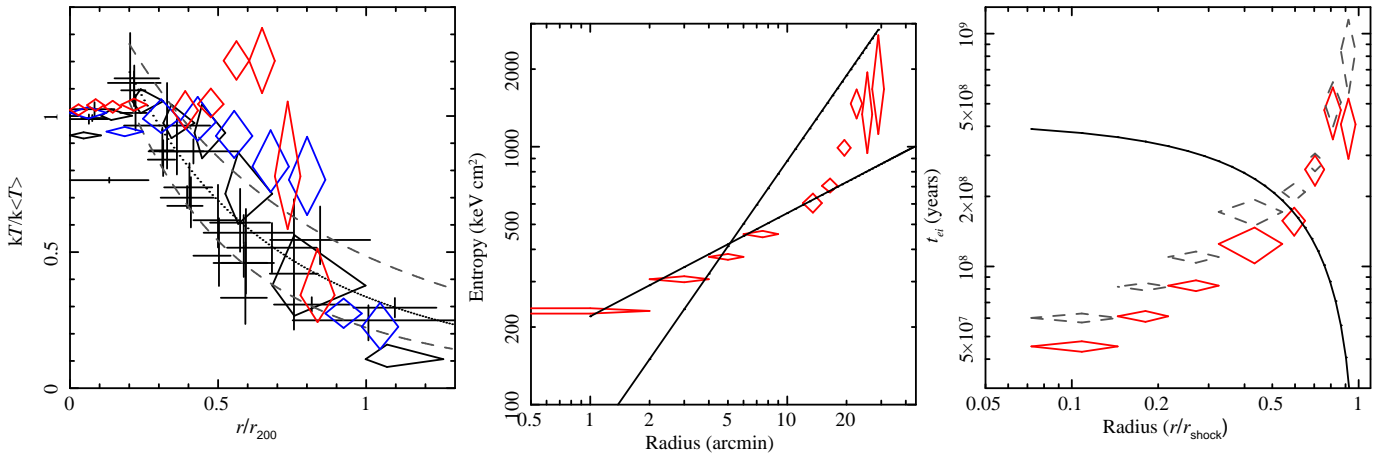

\begin{tabular}{ccc}
\begin{minipage}{0.33333\hsize}
(a) Normalized temperature profiles
\begin{center}
\includegraphics[scale=0.32,angle=-90]{./comp-latest.ps}
\end{center}
\end{minipage}
\begin{minipage}{0.33333\hsize}
(b)  Entropy profiles
\begin{center}
 \includegraphics[scale=0.32,angle=-90]{./entropy.ps}
\end{center}
\end{minipage}
\begin{minipage}{0.33333\hsize}
(c) Equilibrium time scale
\begin{center}
 \includegraphics[scale=0.32,angle=-90]{./eqtime.ps}
\end{center}
\end{minipage}
\end{tabular}
\caption{\label{fig:comp}
Radial profiles of  normalized temperature, entropy and equilibration time scale.  
(a) Scaled projected temperature profiles compared with relaxed clusters~\citep{akamatsu11a}.
The profiles have been normalized to the mean (within 0.4 $r_{200}$) temperature.
The $r_{200}$ value is derived from \citet{henry09}.  
Dotted line shows simulation result~\citep{burns10}. 
Two gray dashed lines show standard deviation.
Black crosses are for the relaxed clusters.
Diamonds show merger clusters (black:~A2142, blue:~A3667, red:~A3376).
In entropy profile (b),  black solid line shows $K \propto r^{1.1}$ given by~\cite{voit03}.
 (c) Ion-electron  equilibration time scale.
 Horizontal scale is normalized by the radius of the radio relic.
 Solid curve shows the time after a shock heating assuming a constant shock speed $v=2000$ km s$^{-1}$.
 Red diamonds show the electron-ion equilibration time $t_{ie}$.
 Gray dashed diamond show the ionization time for $n_{e}t=3\times10^{12} \rm cm^{-3}s$.
}

\end{figure*}

We normalized the radial temperature profile by the average
temperature determined within $0.4 r_{200}$.  Since A~3376 does not
show a strong cool core, this should give a reasonable average
temperature.
The resulting scaled temperature profile is shown in~Fig.\
\ref{fig:comp} along with the previous Suzaku results (black crosses
and diamonds: taken from \cite{akamatsu11a}).  Clearly, there is a
marked difference between the profiles of relaxed and merging
clusters.  The relaxed clusters (black crosses) generally show a
smooth decline from the center to the outer regions, consistent with
the result of hydrodynamic simulations~\citep{burns10}.  Compared with
these profiles, the A~3376 profile shows a large excess within $0.7
r_{200}$ before the sharp drop at $0.7-0.8 r_{200}$.

Numerical simulation of merging clusters by \citet{paul11}shows a shock feature very similar to the one in
  A~3376. This simulation after about 3 Gyr after a major merger shows
  an arc-like high temperature region whose width is about 200 kpc,
  very similar to the observed temperature enhancement in Fig.\
  \ref{fig:comp}.  This suggests that A~3376 is still evolving after
  several Gyrs elapsed from the last major merger.

The $\Lambda$CDM model predicts that clusters grow
 through subcluster mergers and matter accretion from large-scale
 structures.  Those dynamical events should give strong impacts on
 the ICM temperature structure, such as seen in A~3376.  Since the ICM
 density is very low in the cluster outskirts ($n_e =
 10^{-5}-10^{-3}\rm\ cm ^{-3}$), the radiative cooling and conduction
 times are expected to be very long with the cooling time largely
 exceeding the age of the Universe.  We can naturally expect that the
 temperature structure carries important information about the
 history of the cluster growth.  However, as mentioned above, there
 are no clear signs of dynamical evolution in relaxed clusters.  This
 indicates that the observed peculiar temperature structure in the
 merging clusters will be settled to the universal profile in a
 fairly short time scale, much shorter than the Hubble time.  Since
 cooling time is too long, we have to consider other possibilities such as 
 (i) ICM cooling by adiabatic expansion, and  
 (ii) ICM diffusion caused by pressure gradient.
 We will discuss these time scales in section~\ref{sec:timescale}.

A fraction of the merger energy would be channeled
   into the acceleration of ultra-relativistic particles and
   amplification of magnetic fields. However, it is unknown that how
   these processes actually take place and distribute the merger
   energy to other forms in an efficient way.  
The turbulent motion can be detected through line broadening by X-ray microcalorimeters such as with SXS instrument on ASTRO-H \citep{mituda10} and 
Athena, since X-ray calorimeters provide superior energy resolution by a factor of 20--30 better than that with the CCD instruments. 
The wealth of information will open a new window in our understanding of heating and particle acceleration
   caused by cluster merger shocks.  

\subsection{Mach Number}
\label{sec:shock}
Recent studies with XMM-Newton and Suzaku showed sharp temperature
drops at the region of the radio relic in A~3667~\citep{finoguenov10,
  akamatsu11b}.  \citet{finoguenov10} reported a sharp edge in the
surface brightness at the outer boundary of the NW radio relic in
A~3667\@.  \citet{akamatsu11b} reported a significant jump in the ICM
parameters (temperature, surface brightness, electron density, and
pressure) across the same relic.  The derived Mach number is ${\cal M} \sim 2$.

We estimate the Mach number based on the Suzaku data in the same way
as \citet{akamatsu11b}.  The Mach number can be obtained by applying
the Rankine-Hugoniot jump condition, 

\begin{equation}
\frac{T_2}{T_1}=\frac{5{\cal M}^4+14{\cal M}^2-3}{16{\cal M}^2}, ~~
 \frac{1}{C}=\frac{3}{4{\cal M}^2}+\frac{1}{4}, 
 \end{equation}

 where $C=\frac{n_{e1}}{n_{e2}}$ is the shock compression and subscripts 1 and 2 denote pre-shock and
  post-shock values, respectively, assuming the ratio of specific
  heats as $\gamma=5/3$.  Here, we assume the regions
\timeform{21'}-\timeform{24'} and \timeform{27'}-\timeform{31'} to be
the post and pre shock regions, respectively.  
Because the compression factor, $C$, exceeds the value for the strong shock limit,
we only derived mach number from temperature jump.
Table~\ref{tab:mach} shows the resultant Mach number, whose values are ${\cal M}=2.94\pm0.77$ based on the jumps of temperature.  
Compared with other clusters
which also show shock fronts ~(A520, 1E0657-558, A2246, A~3667:
~\cite{markevitch05,clowe06, russell10, akamatsu11b}), the Mach number
in A~3376 is slightly larger than the other cases.

We derive
the compression factor, $C = 7.8 ~\pm~4.7$, which almost exceeds the value for
the strong shock limit, even though with a large error.  The electron
density estimated here is based on the assumption of spherical
symmetry, which is not strictly correct when the shock propagates along the
merger axis.  Since the spherical assumption relates the observed flux
to a shell-like volume with a larger line-of-sight depth than the case
of a relic volume, the density in particular for the pre-shock region
is underestimated by a factor of roughly 1.5.  In addition to this
uncertainty, we checked the dependence on the centroid position of the
spherical shock.  We set the center of the sphere at the secondary
X-ray peak which is closer to the radio relic.  The resultant
compression parameter is $C=5.01$.  This indicates that the
determination of the centroid have a significant impact on the shock
parameter, which is mainly resulted from the estimated line of sight
depth of the shocked region.

The shock front in A~3376 is located far from the cluster center~(1.2
Mpc $\sim~0.6~r_{200}$) just as in A~3667\@.  The temperature in the
upstream region of the shock is $\sim 1\rm ~keV$, indicating the
pre-shock sound speed to be $v_{\rm ss}\sim 520$ km~s$^{-1}$.  
Combining this with the shock compression $C$, we can evaluate
the shock speed by $v_{\rm shock}=C\cdot v_{\rm ss}$ to be
$>2080$ km s$^{-1}$ assuming $C=4.0$ from the above discussion. 
The estimated shock speed well agrees with those in other clusters 
(1E0657-558: 4500 km s$^{-1}$, A520: 2300 km s$^{-1}$), 
but twice higher than the A~3667 case ($1360\pm 120$ km s$^{-1}$).  
The shock speed is large enough to account for the radio relic by a merger shock model \citep{takizawa00,ricker01, mathis05}.  
On the other hand, this $v_{\rm shock}$ seems too high considering the ICM temperature of 4 keV\@.  
We may need to consider a possibility that the true gas temperature is much higher than the measured electron temperature.

\begin{table}[t]
\begin{center}
\caption{Mach number estimation for A~3376 assuming $\gamma=5/3$.} \label{tab:mach}
\begin{tabular}{cccc}\hline \hline
 Parameter & Pre shock & Post shock & Mach number\\ \hline
 $kT$ [keV] & $1.34\pm 0.69$ & $4.68 \pm 0.48$ & $2.97\pm 0.77$ \\
 \hline
\end{tabular}
\end{center}
\end{table}

\subsection{Entropy Profile}
\label{sec:entropy}

Theoretical studies of the ICM~\citep{tozzi01,voit03} predicts the entropy profile approximated by $r^{1.1}$, 
assuming that the gravitational energy of the accreting gas is efficiently converted into the thermal energy.
However, Chandra observations revealed that the
  entropy in the central regions of relaxed clusters often showed a
  flat profile \citep{cavagnolo09}.  The origin of this behavior is
  considered to be the feedback of star formation and central AGN
  activities, which generate extra entropy \citep{peterson06}.
  Entropy profile is thus sensitive to the non-gravitational energy
  injection and can be used to explore such a process even in the
  outer regions of clusters.

  We will define the ``entropy'' of the ICM by $K = kT\cdot
  n_e^{-2/3}$.  Fig.\ \ref{fig:comp} shows the resultant entropy
  profile of A~3376.  The slope shows a large deviation from the
  standard value of 1.1 and close to 0.4 in the radius range
  \timeform{0.5'}-\timeform{20.0'}.  The solid black line shows
  $K\propto r^{0.4}$, compared with the solid gray line indicating
  $K\propto r^{1.1}$.  Recent XMM-Newton study reported entropy
  profiles of 31 clusters \citep{pratt10}, and the distribution of the
  slope indicated a peak value of 0.98.  Therefore, the slope of 0.4
  is very unusual and can be regarded as a transient value occurring
  in a certain phase of cluster mergers.  
 
   The slope shows a steepening and flattening across the radio relic,
   which itself indicates the shock heating.  A similar feature was
   also reported in A~3667 \citep{akamatsu11b}.  The slope of A~3667
   showed a marginal increase and a sudden drop across the radio relic
   region.  In A~3376, the increase of the entropy slope is more
   significant, which suggests a stronger shock than in
   A~3667\@.  This probably corresponds to the feature that the
   estimated Mach number in A~3376 is factor of 2 higher that in
   A~3667\@.

\subsection{Relaxation Time Scales}
\label{sec:timescale}
As shown in the previous sections, A~3376 exhibits
  distinct ICM properties which are significantly different from those
  seen in the relaxed clusters.  The density and temperature
  structures around the shock region, along with the irregular
  morphology of this cluster, suggest that A~3376 is a young system
  regarding the relaxation of the ICM\@.  
  The fact that most of the
  relaxed clusters now attain almost universal temperature profile
  indicates that the time needed for the ICM to reach thermal and
  dynamical equilibrium is much shorter than the cluster life.  In
  this section we evaluate several relevant time scales based on the
  present observational quantities.  

We first estimate the time in which the shock front
  reaches the current position.  Assuming that the shock has traveled
  the distance between the cluster center and the relic position with
  a constant velocity ($v = 2000$ km s$^{-1}$), the time required to
  propagate this length of $\sim 1.5$ Mpc is 0.32 Gyr.  
 
 Next, we estimate the time scale to attain thermal
  equilibrium after the shock heating.
The shock heating should first act on ions rather than electrons, and then the thermal energy is transferred from
ions to electrons through Coulomb collisions.  The equilibration time
scale between ion and electron is 3 orders of magnitude longer than
the electron-electron timescale.  Therefore, the cluster
thermal time scale after the shock heating is limited by the
electron-ion time scale.  According to \citet{takizawa98}, the
electron-ion equilibration timescale is given by
\begin{equation}
t_{\rm ie}=
 2\times 10^{8}{\rm\ yr} \left(\frac{n_{e}}{10^{-3}{\rm\ cm^{-3}}}\right)^{-1}
 \left(\frac{T_{e}}{10^{8}\rm\ K}\right)^{3/2}
 \left(\frac{\ln \Lambda}{40}\right),
\end{equation}
where $\ln \Lambda$ denotes the Coulomb logarithm \citep{spitzer56}.

In the central region, this gives 0.03 Gyr, which is about 10 times
shorter than the elapsed time after the shock passage.  
We calculated the elapsed time after the shock passage
  and the time for electron-ion equilibration $t_{\rm ie}$ as a
  function of radius.  The resultant time scales are shown in Fig.\ \ref{fig:comp} (c).  
  Here, red diamonds show $t_{\rm ie}$, and
  gray dashed diamonds are the ionization time scale  $t_i$ of Fe-K$\alpha$ assuming
  $n_et=3\times10^{12}$ cm$^{-3}$~s.  Solid line shows the elapsed
  time after the shock passage, assuming the shock propagation from
  inner to outer regions.  We can see that the elapsed time is shorter
  than $t_{\rm ie}$ outside of $0.6 r_{\rm shock}$ corresponding to
  $17'$ (920 kpc).  This suggests that the electron temperature is
  likely to be lower than the ion temperature in the outer region
  including the radio relic \citep{akahori08, akahori11}.  
  Those sign already reported another merging cluster RX J1347.5-1145~\citep{ota08}.
  In the region that $t_{i}$ exceed the sound crossing time $t_{sc}$ ( $t_{i} >~t_{sc}$),
  there will be non-equilibrium ionization state.
  The future high resolution X-ray spectroscopy can reveal those dynamical ionization states.
  
Finally, we estimate the time scale in which the
  sharp change of the temperature disappears.  Those high temperature
  region will diffuse out due to the high pressure in the post-shock
  region and reach equilibrium in the sound crossing time
  ($t_{sc}=R/v_{s}$), where $R$ is the width of the high temperature
  region and $v_{s}$ is the sound speed, respectively.  Using the
  observed temperature ($kT= 4.7$ keV which corresponds to
  $v_{s}=1100$~km s$^{-1}$) and the width of the high temperature
  region ($R=320$ kpc), we can calculate the sound crossing time.  The
  resultant time scale is $t_{sc}=0.28$ Gyr.

In summary of this section, the observed features of temperature change, entropy slope and the time scale consideration
  all point to that the outer region ($> 1$ Mpc) of A~3376 is not reaching the equilibrium. 
 In this sense, A~3376 is still a very young system in view of the cluster evolution.  

\section{Summary}
We observed Abell~3376 with Suzaku XIS and derived
  radial profiles of temperature, surface brightness, electron
  density, and thermal pressure.  The temperature and surface brightness
  shows remarkable jumps across the radio relic, located at $\sim 1.5$
  Mpc from the cluster center.   We evaluated the Mach number using
the Rankine-Hugoniot jump condition as ${\cal M}=3.0 $.  The main
results on A~3376 are summarized as follows;

\begin{itemize}
\item The ICM temperature is fairly flat at about 4 keV in the radius
 range from the center to $0.6 r_{200}\sim 1.1$ Mpc, followed by a
 small enhancement near the radio relic to 5 keV\@.
\item Across the radio relic region, the ICM parameters (temperature,
  electron density, and pressure) show significant jumps.  The
  temperature drops from 4.5 keV to 1.0 keV, and the density and
  pressure both show an order of magnitude drop.
\item The estimated Mach number of the shock is ${\cal M}=2.97\pm 0.77$
 based on the temperature jump, leading to the shock speed of $v_{\rm
   shock}> 2000$ km s$^{-1}$.

 \item The temperature structure observed in A~3376 is markedly
   different from those in relaxed clusters, suggesting that this
   cluster is a young system with the ICM properties strongly
   influenced by the merger shock.
 \item The entropy profile is significantly flatter than the standard
   prediction of $r^{1.1}$ and is approximated by $r^{0.4}$,
   accompanied by slope changes across the radio relic. This indicates
   that the heating process is still going on in this region.
 \item Based on the measurement temperature and electron density, the ion-electron
   relaxation time is longer than the elapsed time after the shock
   passage near the shock region.  

 \end{itemize}

These results show that A~3376 is a dramatic
   merging cluster, with their outskirts still under a non-equilibrium
   condition.  Because of the existence of the large scale shock
   front, future X-ray and SZ observations will provide new insight on
   the dynamical evolution of clusters.  

\bigskip The authors thank all the Suzaku team members for their
support of the Suzaku project.  
We also thank an anonymous referee for constructive comments.
H. A. is supported by a Grant-in-Aid for Japan Society for the Promotion of Science (JSPS) 
Fellows (22$\cdot$1582).

\appendix
\section{Stray Light}

To assess the effect of stray light from cluster core to outskirts, we
calculated photon contribution from each region to ``On-Source''
region using {\it xissim} \citep{ishisaki07}.  We use $\beta$-model
($\beta=0.40, r_{c}=\timeform{2.03'}$) surface brightness profiles for
input image and observed spectral information for input spectrum.
Table~\ref{tab:stray} shows the resultant photon fraction detected in
each region.  The ``On-Source'' fraction shows the highest value
followed by those in the adjacent regions, which is naturally
understood considering the point-spread function of the Suzaku XRT
\citep{serlemitsos07}.

\begin{table*}[]
\begin{center}
\caption{Relative count contributions (\%) due to PSF broadening of the Suzaku XIS mirror. Most counts come from the ``On-Source'' region.}
\begin{tabular}{cccccccccccc}\hline
 Detector/ Sky  & (1) & (2) &(3) &(4)& (5)&(6)&(7)& (8)&(9)&(10)\\ \hline
(1) \timeform{0.0'}-\timeform{2.0'} &71.4 & 26.2 & 1.9 & 0.4 & 0.1 & 0.0 & 0.0 & 0.0 & 0.0 & 0.0 \\
(2) \timeform{2.0'}-\timeform{4.0'} &22.3 & 62.6 & 13.5 & 1.3 & 0.2 & 0.0 & 0.0 & 0.0 & 0.0 & 0.0 \\
(3) \timeform{4.0'}-\timeform{6.0'} &3.7 & 25.1 & 58.3 & 12.2 & 0.6 & 0.0 & 0.0 & 0.0 & 0.0 & 0.0 \\
(4) \timeform{6.0'}-\timeform{9.0'} &1.2 & 3.2 & 18.7 & 70.2 & 6.4 & 0.1 & 0.1 & 0.1 & 0.0 & 0.0 \\
 (5) \timeform{12.0'}-\timeform{15.0'} & 0.2 & 0.3 & 0.9 & 10.6 & 76.3 & 10.1 & 1.0 & 0.3 & 0.0 & 0.2 \\
(6) \timeform{15.0'}-\timeform{18.0'} & 0.1 & 0.2 & 0.0 & 0.3 & 11.7 & 70.3 & 15.9 & 1.1 & 0.2 & 0.1 \\
(7) \timeform{18.0'}-\timeform{21.0'} & 0.0 & 0.2 & 0.2 & 0.2 & 1.8 & 13.4 & 68.4 & 14.7 & 0.9 & 0.2 \\
(8) \timeform{21.0'}-\timeform{24.0'}  &0.2 & 0.2 & 0.2 & 0.1 & 0.7 & 1.3 & 17.0 & 65.9 & 13.7 & 0.6 \\
(9) \timeform{24.0'}-\timeform{27.0'}&0.2 & 0.4 & 0.3 & 0.4 & 0.5 & 0.7 & 1.6 & 20.0 & 66.3 & 9.7 \\
(10) \timeform{27.0'}-\timeform{31.0'}	&0.2 & 0.3 & 0.2 & 0.4 & 0.7 & 0.3 & 1.0 & 2.5 & 21.3 & 73.1 \\ \hline
\label{tab:stray}
\end{tabular}
\end{center}
\end{table*}

\section{ICM emission in the outermost region}
Since the validity of our detection of the ICM emission in the outermost region, $r = 27'-31'$, 
is important in this paper, we performed two additional checks. 
One is to look into the Suzaku spectrum by employing different background data, 
and the other is the analysis of the ROSAT pointing data for this cluster.

To examine the adequacy of our background estimation process, 
we incorporated another background data which were taken at 4 degrees offset from A3376 on TX COL (ObsID: 404031010).  
We evaluated the background components (MWH and LHB) in  the same fashion as described in Sec \ref{sec:bgd}.  
The resultant best-fit spectral parameters are shown in Table. \ref{tab:bgd2}.  
Compared with the previous background estimation (Table \ref{tab:bgd}), 
the temperatures and normalizations are consistent within the statistical errors, 
indicating that our background estimation is appropriate.
Using the above background components,
we analyze the outermost region ($27'-31'$ annulus).
The resultant best-fit values are $kT=1.37_{-0.33}^{+0.51}$ keV and norm=$0.3 \pm 0.2$,
which are consistent with the original one.

Next, we derived the surface brightness of A3376 west direction using
the ROSAT archival data (seqID: RP800154N00).  
We set the cluster center at ($\timeform{20h12m31s}, -56^\circ49'12''$) and 
extracted a region with a width of $\timeform{6'}$ to the west from the A3376 center.  
We confirmed that the surface brightness dropped by an order of magnitude from 
$r = \timeform{20'}$ to $\timeform{30'}$ (Fig.6). 
The brightness boundaries defined by the systematic error, mostly due to
the background fluctuation, are indicated with blue histograms. 
Note that the ICM shows significant X-ray emission around $r = \timeform{30'}$ even 
considering the systematic error (rp800154n00\_bk1.fits).
These results indicate the validity of the Suzaku detection of the ICM emission in the outermost region.

\begin{table*}[t]
\caption{ Best-fit background parameters using TX COL}
\begin{center}
\begin{tabular}{lcccccccccccc}
\hline \hline
     			  & $kT (\rm keV)$  &$norm^{\ast}$  ($\times 10^{-3}$)& $kT (\rm keV)$  &$norm^{\ast}$  ($\times 10^{-3}$)& $\chi^{2}$/d.o.f   \\ \hline
NOMINAL	&	   0.08 (fix) 	& $ {9.3}^{+3.7}_{-4.7} $  		& $  {0.29}^{+0.05}_{-0.04}  $ &   $  {5.1}^{+1.9}_{-2.8}  $ &296 / 247 	\\ \hline
\multicolumn{6}{l}{\footnotesize
*:Normalization of the apec component scaled with a factor 1/400$\pi$.}\\
\multicolumn{6}{l}{\footnotesize
Norm=$\rm \frac{1}{400\pi}$$\int n_{e}n_{H} dV/(4\pi(1+z^2)D_{A}^2)\times 10 ^{-14} \rm ~cm^{-5}~arcmin^{-2}$, where $D_A$ is the angular diameter distance to the source.}\\
\end{tabular}
\label{tab:bgd2}
\end{center}
\end{table*}

\begin{figure*}[t]
\begin{center}
\includegraphics[scale=0.42]{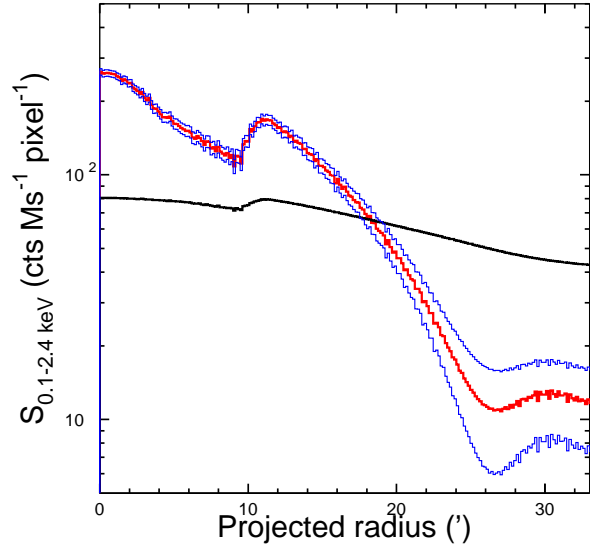}
\end{center}
\caption{
Radial profile of the surface brightness.  
Red line show the ICM signal from A3376. Black line shows the background component.
Two blue lines show the systematic errors considering fluctuations of the background component ($\pm10\%$). 
}
\label{fig:rosat}
\end{figure*}



\begin{thebibliography}{00}
\bibitem[Akahori 
\& Yoshikawa(2011)]{akahori11} Akahori, T., \& Yoshikawa, K.\ 2011, arXiv:1109.0826 

\bibitem[Akahori  \& Yoshikawa(2008)]{akahori08} Akahori, T., \& Yoshikawa, K.\ 2008, \pasj, 60, L19 


\bibitem[Akamatsu et al.(2011a)]{akamatsu11a} Akamatsu, H., Hoshino,
A., Ishisaki, Y., Ohashi, T., Sato, K., Takei, Y.,
\& Ota, N.\ 2011, arXiv:1106.5653

\bibitem[Akamatsu et al.(2011b)]{akamatsu11b} Akamatsu, H., et al.\
2011, arXiv:1111.5162

\bibitem[Ajello et al.(2009)]{ajello09} Ajello, M., Rebusco, P., 
Cappelluti, N., et al.\ 2009, \apj, 690, 367 

\bibitem[Anders \& Grevesse(1989)]{anders89} Anders, E., \&
 Grevesse, N.\ 1989, \gca, 53, 197
 
 \bibitem[Asai et al.(2004)]{asai05} Asai, N., Fukuda, N., 
\& Matsumoto, R.\ 2004, Journal of Korean Astronomical Society, 37, 575 

\bibitem[Bagchi et al.(2006)]{bagchi06} Bagchi, J., Durret, F., 
Neto, G.~B.~L., \& Paul, S.\ 2006, Science, 314, 791 

\bibitem[Bautz et al.(2009)]{bautz09} Bautz, M.~W., et al.\  2009, \pasj, 61, 1117


\bibitem[Burns et al.(2010)]{burns10} Burns, J.~O., Skillman, 
S.~W., \& O'Shea, B.~W.\ 2010, \apj, 721, 1105 

\bibitem[Br{\"u}ggen et al.(2011)]{brueggen11} Br{\"u}ggen, M., 
Bykov, A., Ryu, D., R\"ottgering, H.\ 2011, \ssr, 138 

\bibitem[Cavagnolo et al.(2009)]{cavagnolo09} Cavagnolo, K.~W., 
Donahue, M., Voit, G.~M., \& Sun, M.\ 2009, \apjs, 182, 12 

\bibitem[Clarke 
\& Ensslin(2006)]{clarke06} Clarke, T.~E., \& Ensslin, T.~A.\ 2006, \aj, 131, 2900 

\bibitem[Clowe et al.(2006)]{clowe06} Clowe, D., Brada{\v c},
M., Gonzalez, A.~H., Markevitch, M., Randall, S.~W., Jones, C.,
\& Zaritsky, D.\ 2006, \apjl, 648, L109

\bibitem[Dickey \& Lockman(1990)]{dickey90} Dickey, J.~M., \& Lockman,
 F.~J.\ 1990, \araa, 28, 215

\bibitem[Ebeling et al.(1996)]{ebeling96} Ebeling, H., Voges, W., 
Bohringer, H., et al.\ 1996, \mnras, 281, 799 

\bibitem[George et al.(2008)]{george08} George,M.R.,Fabian, A. C.,
  Sanders,J. S., Young., A. J., and Russell, H. R., \ 2008, MNRAS,
  395, 657

\bibitem[Ishisaki et al.(2007)]{ishisaki07} Ishisaki, Y., et al.\
 2007, \pasj, 59, 113

\bibitem[Henry et al.(2009)]{henry09} Henry, J.~P., Evrard,
A.~E., Hoekstra, H., Babul, A., \& Mahdavi, A.\ 2009, \apj, 691, 1307

\bibitem[Hoshino et al.(2010)]{hoshino10} Hoshino, A., et al.\
2010, \pasj, 62, 371

\bibitem[Ferrari et al.(2008)]{ferrari08} Ferrari, C., Govoni,  F., Schindler, S., Bykov, A.~M., \& Rephaeli, Y.\ 2008, \ssr, 134, 93 

\bibitem[Finoguenov et al.(2010)]{finoguenov10} Finoguenov, A.,
Sarazin, C.~L., Nakazawa, K., Wik, D.~R.,
\& Clarke, T.~E.\ 2010, apj, 715, 1143



\bibitem[Kaastra et al.(1996)]{kaastra96} Kaastra, J.~S., Mewe,
R., \& Nieuwenhuijzen, H.\ 1996, UV and X-ray Spectroscopy of Astrophysical and Laboratory Plasmas, 411


\bibitem[Kawaharada et al.(2010)]{kawaharada10} Kawaharada, M., et  al.\ 2010, \apj, 714, 423


\bibitem[Kawano et al.(2009)]{kawano09} Kawano, N., et al.\ 
2009, \pasj, 61, 377 

\bibitem[Kushino et al.(2002)]{kushino02} Kushino, A., Ishisaki,
Y., Morita, U., Yamasaki, N.~Y., Ishida, M., Ohashi, T.,
\& Ueda, Y.\ 2002, \pasj, 54, 327


\bibitem[Markevitch et al.(2005)]{markevitch05} Markevitch, M.,
Govoni, F., Brunetti, G., \& Jerius, D.\ 2005, \apj, 627, 733

\bibitem[Mathis et al.(2005)]{mathis05} Mathis, H., Lavaux, G.,
Diego, J.~M., \& Silk, J.\ 2005, \mnras, 357, 801

\bibitem[Mitsuda et al.(2010)]{mituda10} Mitsuda, K., et al.\ 2010, \procspie, 7732,  

\bibitem[Nevalainen et al.(2004)]{nevalainen04} Nevalainen, J., 
Oosterbroek, T., Bonamente, M., \& Colafrancesco, S.\ 2004, \apj, 608, 166 

\bibitem[Ota et al.(2008)]{ota08} Ota, N., Murase, K., Kitayama, T., et al.\ 2008, \aap, 491, 363 



\bibitem[Parrish et al.(2011)]{parrish11} Parrish, I.~J., 
McCourt, M., Quataert, E., \& Sharma, P.\ 2011, arXiv:1109.1285 

\bibitem[Paul et al.(2011)]{paul11} Paul, S., Iapichino, L., 
Miniati, F., Bagchi, J., \& Mannheim, K.\ 2011, \apj, 726, 17 

\bibitem[Pratt et
al.(2010)]{pratt10} Pratt, G.~W., et al.\ 2010, \aap, 511, A85

\bibitem[Peterson 
\& Fabian(2006)]{peterson06} Peterson, J.~R., \& Fabian, A.~C.\ 2006, \physrep, 427, 1


\bibitem[Reiprich et al.(2009)]{reiprich09} Reiprich, T.~H., et al.\
  2009, \aap, 501, 899

\bibitem[Ricker \& Sarazin(2001)]{ricker01} Ricker, P.~M., \& Sarazin, C.~L.\ 2001, \apj, 561, 621 

 \bibitem[Russell et al.(2010)]{russell10} Russell, H.~R., 
Sanders, J.~S., Fabian, A.~C., Baum, S.~A., Donahue, M., Edge, A.~C., 
McNamara, B.~R., \& O'Dea, C.~P.\ 2010, \mnras, 406, 1721 
 
\bibitem[Takei et al.(2011)]{takei11} Takei, Y., Akamatsu, H. et
al.\ 2011, Suzaku Conference


\bibitem[Tawa et al.(2008)]{tawa08} Tawa, N., et al.\ 2008,
\pasj, 60, 11

\bibitem[Serlemitsos et al.(2007)]{serlemitsos07} Serlemitsos, P.~J.,
 et al.\ 2007, \pasj, 59, 9

\bibitem[Simionescu et al.(2011)]{simionescu11} Simionescu, A., et al.\ 2011, Science, 331, 1576

\bibitem[Sugawara et al.(2009)]{sugawara09} Sugawara, C., 
Takizawa, M., \& Nakazawa, K.\ 2009, \pasj, 61, 1293 

\bibitem[{{Spitzer}(1956)}]{spitzer56} {Spitzer},L.Jr., 1956,
Physics of Fully Ionized Gases, New York: Interscience

\bibitem[{{Takizawa}(1998)}]{takizawa98} {Takizawa},M., 1998, \apj, 509, 579

\bibitem[Takizawa(2008)]{takizawa08} Takizawa, M.\ 2008, \apj, 687,
  951

\bibitem[Takizawa 
\& Naito(2000)]{takizawa00} Takizawa, M., \& Naito, T.\ 2000, \apj, 535, 586 

\bibitem[Tozzi 
\& Norman(2001)]{tozzi01} Tozzi, P., \& Norman, C.\ 2001, \apj, 546, 63 

\bibitem[van Weeren et al.(2010)]{weeren10} van Weeren, R.~J.,
R{\"o}ttgering, H.~J.~A., Br{\"u}ggen, M.,
\& Hoeft, M.\ 2010, Science, 330, 347

\bibitem[Voit et al.(2003)]{voit03} Voit, G.~M., Balogh,
M.~L., Bower, R.~G., Lacey, C.~G., \& Bryan, G.~L.\ 2003, apj, 593, 272

\bibitem[Wik et al.(2009)]{wik09} Wik, D.~R., Sarazin, C.~L., 
Finoguenov, A., et al.\ 2009, \apj, 696, 1700 




\end{thebibliography}
\end{document}